\documentclass[12pt]{article}

\usepackage{graphicx}  
\usepackage{amssymb}   

\usepackage{caption}    
\title{Econometrics of Insurance  \\ with Multidimensional Types}
\author{G. Aryal, I. Perrigne, Q. Vuong \& H. Xu}
\date{}

\newcommand{\Real}{\hbox{\it I\hskip -2pt R}}
\newcommand{\Unit}{\hbox{\it 1\hskip -3pt I}}

\newcommand{\Natural}{\hbox{\it I\hskip -2.5pt N}}

\begin{document}
\begin{titlepage}
\LARGE

\begin{center}
Econometrics of Insurance \\ with Multidimensional Types\\

\large
\vspace{0.15in}
Gaurab Aryal\\
Boston University\\
\vspace{0.1in}

Isabelle Perrigne\\
Rice University\\
\vspace{0.1in}

Quang Vuong\\
New York  University\\

\vspace{0.1in}

Haiqing Xu\\
University of Texas Austin

\vspace{0.3in}
September 2024\\
\end{center}
\normalsize

\vspace{0.2in}
\normalsize
\noindent
This paper draws from Aryal, Perrigne and Vuong (2016).  It extends it in important directions while including a new section on estimation.
We  thank  Pierre-Andr\'e Chiappori, Liran Einav, Matt Shum and Ken Wolpin as well as two referees and the Editor  for constructive comments. 
We also benefited from participants' comments at  the Stanford Institute for Theoretical Economics,  North American Meeting of the Econometric Society and  seminars at   the Australian National University, Collegio Carlo Alberto,  Columbia University,  Georgetown University, London School of Economics,  National University of Singapore, Paris School of Economics, Sciences Politiques-Paris, Stanford University, University of Fortaleza, University of Pennsylvania,  University of Sydney and University of Wisconsin at  Madison.   
Isabelle Perrigne and Quang Vuong gratefully acknowledge financial support from the National Science Foundation through  grant SES 1148149.

\smallskip\noindent
Correspondence to   Isabelle Perrigne: iperrigne@gmail.com
  
\normalsize
\end{titlepage}

\newpage
\thispagestyle{empty}

\begin{abstract}

In this paper, we address the identification and estimation of insurance models where insurees have private information about their risk and risk aversion. The model  includes random damages and allows for several claims, while  insurees choose from a finite number of coverages.  We show that the joint distribution of risk and risk aversion is nonparametrically identified despite bunching due to multidimensional types and a finite number of coverages. Our identification strategy  exploits the observed number of claims
as well as an exclusion restriction, and a full support assumption.  Furthermore,  our results apply to any form of competition.  We propose a novel estimation procedure combining nonparametric estimators  and GMM estimation that we illustrate in a Monte Carlo study.

\end{abstract}

Keywords:  Insurance, Identification, Nonparametric Estimation,  Multidimensional Adverse Selection, Risk Aversion.

\setcounter{page}{0}

\begin{center}
\maketitle
\end{center}

\section{Introduction}

Insurance is a long-studied topic in economics and is at the core of recent empirical research. Rothschild and Stiglitz (1976) and Stiglitz (1977) provide  benchmark models of insurance under private information on insurees' risk. In  empirical studies, testing adverse selection in risk has generated a large number of papers with mixed results.\footnote{See Chiappori and Salani\'e (2000)
for the most well known test and Cohen and Siegelman (2010) for a survey of  empirical findings.} The empirical literature shows that insurance markets involve heterogeneity in both risk and risk aversion.
See, e.g., Finkelstein and McGarry (2006) for long-term care insurance, 
Cohen and Einav (2007) for automobile insurance,  and
Fang, Keane and Silverman (2008) for health insurance.\footnote{See also 
Cutler, Finkelstein and McGarry (2008) and Einav and Finkelstein (2011) for surveys.}  As noted in these papers, heterogeneity in risk aversion may overturn the prediction of the benchmark adverse selection model that risk and insurance coverage  have a positive correlation. For instance,
a low-risk individual may buy  higher coverage because of high risk aversion and conversely. Whether risk aversion or risk is the primary determinant of the demand for insurance has distinct welfare implications and policy recommendations. Thus, a model of insurance needs  to incorporate insurees' heterogeneity in risk aversion resulting in multidimensional screening and pooling. 

Letting each  insuree be characterized by two parameters capturing his/her risk preference and  risk, 
this paper addresses the nonparametric  identification  and estimation of the joint distribution of risk and risk aversion  from a finite number of  coverage choices with random damages and  multiple claims.  
Allowing for a flexible dependence between risk and risk aversion is important for policy recommendations. Moreover,
our identification result  requires minimal  assumptions on the supply side. In particular, it does not rely either on an insurer's model of coverage offering or on how insurers compete, thereby avoiding 
well-known  complexities of optimal contracting with multidimensional types. See e.g. Rochet and Chone (1998)  and Rochet and Stole (2003).
Identification is a key step for the econometric and empirical analysis of structural models. 
First, it highlights which variations in the data allow one to identify  model primitives.  Second, our identification argument is constructive. It provides the basis for our proposed  estimation method.

We consider a finite number of automobile  coverages of the form  `premium and deductible' although  
our results  apply  to other insurance markets and/or other types of coverages as discussed later.
Since there is no  one-to-one  mapping between 
 the deductible and the insuree's private information due to multidimensional types (risk and risk aversion) and a finite number of coverages, the number of claims plays a key role in identifying the marginal distribution of risk.  To  identify the joint distribution of risk and risk aversion, we exploit an exclusion restriction and a support assumption  that requires sufficient variations in some exogenous characteristics.

The previous results are derived with two offered coverages of the form `premium and deductible' under the widely used specifications of  a Constant Absolute Risk Aversion (CARA)
utility function and a Poisson distribution for the number of accidents. The CARA parameter and the mean of the Poisson distribution then measure the risk preference and risk for each insuree, respectively.  We show that our identification results extend to more than two offered coverages and  a larger class  of models beyond the CARA attitude toward risk and the
Poisson distribution for the number of accidents. Observing more coverages helps identify the joint distribution of risk and risk aversion by alleviating the full support assumption.
Moreover, our results also extend to health insurance coverages of the form `premium, deductible, and copayment' with a fixed or proportional deductible.
Regarding  estimation,  we develop a new multi-step  procedure that is computationally friendly. The procedure combines several nonparametric estimators  (kernel and sieves estimators) and GMM estimation.  A Monte Carlo study illustrates our estimation procedure.

Our paper differs on several aspects from the previous literature on the identification and estimation of models under incomplete information. 
Multidimensional adverse selection leads to bunching or pooling,  making model identification a challenging problem. In particular,  identification can no longer rely on the one-to-one  equilibrium mapping(s) between the agent's unobserved continuous types and his observed 
outcome(s)/action(s).\footnote{See Luo, Perrigne and Vuong (2017) who study identification of nonlinear pricing models with multiple types relying on  Armstrong (1996) model and Aryal and Zincenko  (2023)  who study identification of Rochet and Chone (1998) model. In contrast,  Kong, Perrigne and Vuong (2022)  exploit the bidders' multiattribute bids in auctions to avoid the complexity of  the optimal mechanism.}
A finite number of contracts often leads to pooling with similar identification issues.\footnote{Crawford and Shum (2007)  and Gayle and Miller (2015)  circumvent this issue by considering as many  contracts as agents'  (one dimensional)  discrete types.}  
We develop a different identification strategy relying on the insurees' choice of coverage and the observed number of claims.
Given that we  do not rely on the optimality of the offered coverages, our results apply to any form of competition in the insurance industry. 
This result contrasts with the previous literature and provides a novel perspective on the identification of models under incomplete information.  In particular,  the estimation of the model primitives can no longer rely on  inversion as in Guerre, Perrigne and Vuong (2000) or quantiles as in Marmer and Shneyerov (2012) or Luo, Perrigne and Vuong (2018). As a consequence, our paper proposes a new estimation procedure.

The  paper is organized as follows. Section 2 presents the model, whereas Section 3 studies its identification and an important extension  with more than two contracts.  Section 4 presents our new estimation procedure and a Monte Carlo study.
Section 5 concludes. An online appendix
 collects some  proofs, auxiliary results and several extensions.  See Aryal, Perrigne, Vuong and Xu (2024).

\section{A  Model of Insurance}

This section presents a model in which insurees have private information about their  risk and risk aversion when buying insurance from a finite number of coverages. In the presence of multivariate private information and/or a finite number of coverages, pooling arises as individuals of different types choose the same coverage.   To fix ideas and  in the spirit of the early literature on adverse selection, we consider automobile insurance  throughout the paper although our framework  also applies to (say) homeowner and rental insurance. See the  Appendix for a discussion of an application to  health insurance.

\bigskip\noindent
{\sc Motivation}

Our model draws from  Stiglitz (1977), wherein  insurees are heterogeneous in their  probability of accident (risk), which is their private information,  but are homogeneous in  risk aversion.  However,  Finkelstein and McGarry (2006) and  Cohen and Einav (2007) find that  heterogeneity in risk aversion might be more important than  heterogeneity in risk across insurees.  Thus, we consider that  risk aversion is as heterogeneous and private  like the probability of an accident.  Consequently, asymmetric information becomes bidimensional.
In addition, Stiglitz (1977) assumes that there can be at most one accident with  fixed damage. In reality, there might be more than one accident during the policy period, and ex-ante, every accident involves  random damage.

 Ignoring this bidimensional feature may affect insurance policy design.
For instance,  an insuree with a low probability of an accident and a high risk aversion may buy a contract with  high coverage, i.e., a low deductible.
Similarly, an insuree with a high probability of an accident and a low  risk aversion may buy a contract with  low coverage, i.e., a high deductible.  These two examples contrast with Stiglitz (1977) predictions as the former should choose a low coverage and the latter a high coverage when insurees have homogeneous preferences.  Furthermore, when risk and risk aversion are negatively correlated,  this leads to  advantageous selection where high coverages are bought by insurees with low risk  but high risk aversion.   Using a  probit regression for the choice of deductible and a Poisson regression for the number of claims on a set of insurees' characteristics,  Cohen and Einav (2007) show,   for instance,  that   married,  educated  and female insurees  tend to have fewer accidents while buying a high coverage.  Their structural analysis  confirms that these insurees tend to be more risk averse.
 It is, therefore, crucial to consider heterogeneity in risk aversion.
Similarly, the distribution of damages  may have an important effect on the insuree's choice of coverage.  
For instance, at a given level of risk and risk aversion, a higher expected damage will induce the insuree to choose more coverage.

 In view of this, our model includes multiple accidents with random damages and heterogeneity in  risk and   risk aversion. 
Although insurance contracts may take various forms,  we  consider the benchmark case of premium-deductible contracts. Our results, however,  extend to other forms of insurance coverage.  For instance the Appendix contains an extension to health insurance with a copayment per claim  in addition to a premium and a deductible for the  coverage period.

\bigskip\noindent
{\sc  Model Assumptions}

We make the following assumptions. In our model, the insuree's risk
$\theta$ is  the expected number of accidents during the coverage period  whereas the parameter $a$ measures the insuree's risk aversion. They satisfy the following assumptions.

\medskip\noindent
{\bf Assumption A1:}{\em

\noindent
(i) An insuree's utility function exhibits Constant Absolute Risk Aversion (CARA),  i.e., $U(x;a)= - \exp(-ax)$ where $a>0$,

\noindent
(ii) The types $(\theta,a)$  are jointly distributed as $F(\cdot,\cdot)$  with positive density $f(\cdot,\cdot)$ on its support 
 $\Theta \times {\cal A}= (\underline{\theta},\overline{\theta}) \times (\underline{a},\overline{a}) \subset \Real_{++} \times \Real_{++}$, 

\noindent
(iii)  Conditional on $\theta$,   the number  $J$ of accidents each insuree may have follows a 
Poisson distribution ${\cal P}(\theta)$, i.e., $p_j(\theta)={\rm Pr}[J=j|\theta]= e^{-\theta} \theta^j/j!$ for $j=0,1,\ldots$, 

\noindent
(iv)  Each accident involves a damage $D_j$ independently distributed as $H(\cdot)$ on $(0,\overline{d}) \subset \Real_+$ for $j=1,\ldots,J$.

}

\medskip
Individual and car characteristics  are introduced later in Assumption A2.
By A1-(i), the utility function is increasing and concave. The CARA specification has two main advantages: (i) it keeps the model tractable, and (ii) the attitude toward risk is independent of initial wealth. In most empirical settings, the agent's wealth is not observed, making the CARA utility specification quite convenient. These properties have made the CARA utility  popular  in the theoretical and empirical literature.
By A1-(ii), each insuree is characterized by a pair $(\theta,a)$, which is his/her private information. 
Assumption A1-(iii) specifies the distribution of accidents as Poisson with mean $\theta$. This distribution is widely used in actuarial science to model 
the number of accidents for a given individual.
The distribution  $F(\theta)$ of risk is left unspecified, so the distribution of the number of accidents in the population is a flexible nonparametric mixture of Poisson distributions, namely ${\rm Pr}(J=j)= \int_{\Theta} p_j(\theta) dF(\theta)$.\footnote{Using a mixture of Poisson distributions to model the number of accidents in a population dates back to Greenwood and Yule (1920) where the mixing distribution is a Gamma distribution thereby leading to a Negative Binomial distribution for the population.  Cohen and Einav (2007) consider a  log normal  mixture of Poisson  distributions.} 

The combination of the CARA utility and the Poisson distribution is convenient to model  the expected utility and the associated certainty equivalent.
Relaxing the CARA and/or Poisson specifications is possible at the cost of obtaining implicit expressions for the expected utility. 
We consider  alternative specifications for  the agent's utility function and the distribution of accidents in the online appendix.
Assumption A1-(iii,iv)  imply that the number  $J$ of accidents and corresponding damages are independent of $a$ and $(\theta,a)$, respectively.  These independence assumptions will  be relaxed in Section 3 by introducing the insurees' characteristics, such as their  driving experience,  which then allows  for unconditional dependence between the number  $J$ of accidents with $a$ and damages $D_j$ with $(\theta,a)$. 

The model primitives are  the joint distribution of risk and risk aversion  and the damage distribution, i.e.,  $[F(\cdot,\cdot), H(\cdot)]$.
Each insuree is characterized by a pair of types $(\theta,a)$ from $F(\cdot,\cdot)$, which is left unspecified. The insuree chooses among a finite menu of  insurance contracts of the form $[t,dd]$, where $dd$ is the deductible per accident. The insuree chooses the contract that maximizes his expected utility and pays the corresponding premium $t$.
In case of an accident with damage below the deductible, the insuree pays for it.
Otherwise, the insurer pays the damage above the deductible, and the insuree pays the deductible.  

\bigskip\noindent
{\sc Insuree's Choice of Coverages}

Let $C$ be the number of available contracts. To simplify the presentation, we consider   $C=2$. Extending the analysis to $C > 2$ is presented in Section 3. 
Let $(t_1,dd_1)$ and $(t_2,dd_2)$  denote two coverages with $0 < t_1< t_2$ and $\overline{d}>dd_1>dd_2 \geq 0$ so that no contract dominates the other. 
This ordering is the only requirement  we make on the observed coverages.  It is related to rational  offering and choice. Otherwise,  we would have $t_1 < t_2$ and $dd_1 \leq dd_2$ making contract 1 the natural choice for insurees.
These coverage terms
 do not need  to satisfy  profit maximizing conditions for the insurer(s) allowing us to be agnostic about the market structure of the insurance industry. We note that our setting can include  full coverage, which corresponds to $dd_2=0$.
Moreover, the highest deductible $dd_1$ should be smaller than the maximum damage $\overline{d}$ to rationalize buying some insurance.
Indeed, the no insurance option, which corresponds to  $(t_0,dd_0)=(0,\overline{d})$, would dominate $(t_1,dd_1)$  if $dd_1=\overline{d}$ since $t_1>0$.

For a $(\theta,a)$-individual with wealth $w$, his  expected utility with coverage $(t,dd)$ for $t\geq 0$ and $0 \leq dd \leq \overline{d}$ is $V(t, dd;\theta,a,w) \equiv {\rm E}\left[U\left(w-t-\sum_{j=0}^J \min\{dd,D_j\};a\right)| \theta\right]$ where $D_0 \equiv 0$ by convention.  Under A1, we obtain
\begin{eqnarray}
V(t, dd;\theta,a,w)&=& p_0(\theta) U(w-t;a)+p_1(\theta){\rm E}[U(w-t-\min\{dd,D_1\};a)]   \nonumber \\
&&+ p_2(\theta) {\rm E}[U(w-t- \min\{dd,D_1\}-
\min\{dd,D_2\};a)]+\ldots \nonumber\\
                 &=& -p_0(\theta) e^{-a(w-t)}-p_1(\theta)e^{-a(w-t)}{\rm E}[e^{a\min\{dd,D_1\}}]  \nonumber\\
&&-p_2(\theta) e^{-a(w-t)} {\rm E}[e^{a\min\{dd,D_1\}}]{\rm E}[e^{a\min\{dd,D_2\}}]-\ldots\nonumber\\
                 &=&-e^{-a(w-t)}\left[p_0(\theta)+p_1(\theta)\phi_a(dd) +p_2(\theta) \phi_a^2(dd)+\ldots \right]\nonumber\\
                  &=& -e^{-a(w-t)} e^{-\theta}\left( 1 + \frac{\theta \phi_a(dd)}{1!} + \frac{\theta^2\phi_a^2(dd)}{2!}+\ldots\right)\nonumber\\
                  &=& -e^{-a(w-t)+\theta[\phi_a(dd)-1]},
\end{eqnarray}
with $\phi_a(dd)\equiv{\rm E}[e^{a\min \{dd,D\}}]<\infty$ where the expectation is with respect to the random damage $D$.  In particular, $\phi_a(dd) \geq 1$ with equality only if $dd=0$, as $a>0$.
The expression $\phi_a(dd)$ can be interpreted as the expected loss in utils of an accident with  deductible $dd$ for an individual with risk aversion $a$.
 The first equality in (1) considers all the possible  number of accidents and their respective costs for a $(\theta,a)$-individual buying insurance $(t,dd)$.   The second equality uses the CARA utility function  and  
A1-(i,iv).
The third equality uses that damages are identically distributed by A1-(iv).
Lastly, the fourth equality relies on  the Poisson distribution of accidents by A1-(iii).

The $(\theta,a)$-individual chooses the contract that maximizes his expected utility or equivalently his certainty equivalent. 
The certainty equivalent  $CE(t,dd;\theta,a,w)$ of  insurance coverage  is defined as the amount of certain wealth for the insuree that will give him/her the same level of utility when he/she has  coverage, i.e.,  $-\exp(-aCE(t,dd;\theta,a,$ $w))=V(t,dd;\theta,a,w)$. Thus, by (1) we have for $t\geq 0$ and $0 \leq dd \leq \overline{d}$
\begin{eqnarray}
CE(t,dd;\theta,a,w)=w-t-\frac{\theta[\phi_a(dd)-1]}{a}.
\end{eqnarray}
When comparing different insurance coverages for a $(\theta,a)$-individual, i.e., their certainty equivalents,  the individual wealth $w$ then cancels out, which makes the choice of the CARA and Poisson specifications quite convenient. Therefore, wealth need not be observed.
The next lemma establishes the monotonicity in $(\theta,a)$ of the certainty equivalent as well as the frontier  that partitions the $\Theta \times {\cal A}$ space into two subsets ${\cal C}_1$ and ${\cal C}_2$ of individuals choosing coverages 1 and 2, respectively.
 
\medskip\noindent
{\bf Lemma 1:} {\em Let A1 hold.\\
(i) When $dd=0$ (full coverage), the certainty equivalent (2) reduces to $w-t$. When $dd>0$, the certainty equivalent  (2) decreases in both risk and risk aversion. \\
(ii) The frontier separating ${\cal C}_1$ and ${\cal C}_2$ is given by
\begin{eqnarray} 
\theta(a) &=& \frac{a(t_2-t_1)}{\phi_a(dd_1)-\phi_a(dd_2)} 
=    \frac{t_2-t_1}{ \int_{dd_2}^{dd_1} e^{aD} [1-H(D)] dD},
\end{eqnarray}
which is decreasing in $a$. Every $(\theta,a)$-individual below (resp. above) this frontier prefers coverage 1 to coverage 2 (resp. 2 to 1).}\footnote{As a matter of fact, Lemma 1 also applies when $(t_1,d_1)=(0,\overline{d})$ (no insurance), and $(t_2,dd_2)=(t,dd)$ with $t>0$ and $0\leq dd <\overline{d}$.}

\medskip\noindent
The proof is given in the  Appendix.  The first part of (i) is expected as wealth is reduced only by the premium with full coverage.    The second part of (i) is also intuitive as the certainty equivalent, and the utility move together.
Regarding (ii),  the frontier between ${\cal C}_1$ and ${\cal C}_2$ is the locus of $(\theta,a)$-insurees who are indifferent between the two contracts,
i.e., for whom $CE(t_1,dd_1;\theta,a,w)= CE(t_2,dd_2;\theta,a,w)$. 
This frontier is independent of wealth $w$. The denominator of (3)  is  the difference in expected  utility losses from an accident between the two coverages for an individual with risk aversion $a$.

Figure 1 illustrates  the choice between two coverages with a Uniform damage distribution on $[0,10^4]$. In agreement with Cohen and Einav (2007), the range of the CARA parameter $a$ is $[10^{-4}, 10^{-3}]$.  The range of the parameter $\theta$ is $[0.1,1]$. The two coverages are $(t_1,dd_1)=(600,1000)$ and 
$(t_2,dd_2)=(850,500)$. 
The bold curve represents the frontier between the two coverages following (3) with individuals above it preferring coverage 2 to coverage 1. Considering a point on this frontier (say) $(\theta,a)=(0.371,0.005)$, Figure 1 also displays the certainty equivalent isocurves for coverages 1 and 2 in dashed and dotted curves, respectively.  These certainty equivalents decrease as $\theta$ or $a$ increases.\footnote{The top north-east bold curve labelled `Frontier 2 vs 3' can be omitted at this time and is discussed in Section 3.}
For completeness, Figure 1 displays a bold dashed curve in the southwest corner corresponding to the frontier between no insurance and coverage 1, with individuals above it preferring coverage 1.  This frontier is obtained from (3), by letting  the no insurance corresponds  to a zero premium and a deductible at $\overline{d}=10^4$, i.e., to $(t_0,dd_0)=(0,10^4)$, where $\overline{d}$ represents the car value. 

We remark that there is no exclusion since all individuals are willing to buy  insurance because coverage 1 always dominates no insurance, i.e., $CE(t_1,dd_1;\theta,a) \geq CE(0,\overline{d};\theta,a)$ for all $(\theta,a)$ since the individual rationality constraints are always satisfied. Figure 1 can be interpreted as an illustration that the insurance industry must serve everyone, i.e., it  is subject to a universal service requirement. In contrast, if the premium for coverage 1 goes above 619,  the bold dashed curve will shift upward  above  the $(\underline{a},\underline{\theta})$ point so that individuals below this curve would prefer no coverage and are therefore excluded.
If insurance is mandatory, individuals below the corresponding  bold curve (frontier between the two coverages) are forced to buy insurance and will choose coverage 1.

\section{Identification}

In this section, we study the identification of  the joint distribution of risk and risk aversion $F(\theta,a)$ and the damage distribution $H(\cdot)$.  Although  we do not impose any parameterization on $F(\cdot,\cdot)$ or $H(\cdot)$,
 our identification analysis is semiparametric since the insurees' utility function and their probability of accidents are parameterized by $a$ and $\theta$, respectively. We discuss in the online appendix how to relax the CARA and the Poisson specifications to other parameterizations.
Our identification analysis  shows the key role played by the number of accidents, an exclusion restriction and a support assumption. The  identification problem is to recover uniquely the distributions $F(\cdot,\cdot)$ and   $H(\cdot)$ from observables.
We observe  the contract $(t,dd)$ purchased by each insuree, as well as all
his/her $J$ claims with  corresponding  damages $(D_1,\ldots,D_J)$.\footnote{We abstract away from the truncation issue and assume that all the accidents and damages are observed. In the online appendix, we consider the case when only $J^*$ claims with their corresponding damages $(D_1,\ldots,D_{J^*}$ are observed due to the truncation at the deductible $dd$. See Aryal, Perrigne, Vuong and Xu (2024).}  Hereafter,   insurance is mandatory as when  the insurer is subject to universal service. Otherwise, our identification results hold conditional on buying insurance. We first  consider the case of two offered coverages and discuss later the benefits of having more coverages.

\bigskip\noindent
{\sc Introducing  Covariates}

We  introduce   some  observed variables $X$ characterizing the insuree and his/her car. Variables related to the insuree 
may contain age, gender, education, marital status, location and driving experience.  Variables 
related to the car may include  car mileage, business use, car value, power, model, and make. 
The  structure becomes $[F(\theta,a|X), H(D|X)]$ as such variables  may affect 
the insuree's risk and risk aversion as well as damages.
For instance, damages  with  an  expensive car are  likely  larger than those with an inexpensive one, ceteris paribus.   The next assumption specifies the data-generating process. It maintains the CARA and Poisson specifications in A1-(i, iii) while extending A1-(ii, iv) to allow for the characteristics $X$.

\medskip\noindent
{\bf Assumption A2:} {\em  The tuplets $(\theta,a,X,J,D_1,\ldots,D_J)$ are i.i.d. across individuals, and \\
(i) CARA utility function  as in  A1-(i),\\
(ii) $(\theta,a)|X \!\sim\! F(\cdot,\cdot|X)$ with positive density $f(\cdot,\cdot|X)$ on its support $\Theta(X)\! \times\! {\cal A}(X)\!=\!(\underline{\theta}(X),\overline{\theta}(X))\!\times\!(\underline{a}(X),\overline{a}(X))\subset \Real_{++} \!\times\! \Real_{++}$,\\
(iii)  Poisson distribution for the number $J$ of accidents as in  A1-(iii), \\ 
(iv)  The damages $D_j, j=1,\ldots,J$ are i.i.d. as $H(\cdot|X)$ on $(0,\overline{d}(X))\subset \Real_+$.
}

\medskip\noindent
Assumption A2 parallels A1.  As noted,  A2-(iii) states that the number of accidents  $J$ depends only on  the insuree's risk $\theta$ through the Poisson distribution, with $\theta$ being the expected number of accidents.  It implies that $J$ is independent of $(a,X)$ given $\theta$, i.e.,  $J \perp (X,a)\big{|} \theta$.  Similarly, A2-(iv) implies that damages are independent of $(\theta,a)$ given $(J,X)$, i.e., $(D_1,\ldots,D_J) \perp (\theta,a)  \big{|}(J,X)$.  It should be  noted that  utilities and number of accidents indirectly depend on the insuree/car characteristics $X$ through $a$ and $\theta$ which depend on $X$ by A2-(ii). In particular, A2 allows for {\em unconditional} correlations between the number  $J$ of accidents with insurees's risk aversion $a$ and between damages $D_j$ and insuree's types $(\theta,a)$.

The offered coverages may also depend on the vector of characteristics $X$
as $(t_1(X),$  $dd_1(X))$ and $(t_2(X), dd_2(X))$ with $0 < t_1(X) < t_2(X)$ and 
$\overline{d}(X) > dd_1(X) > dd_2(X) \geq 0$.
Following insurance regulations, insurers may not be allowed to use some of the  insurees' observed characteristics as discriminatory tools in the coverage terms 
$(t,dd)$. Thus, $(t,dd)$ may not depend on all the $X$ variables.
Hereafter, we let ${\cal S}_A$ denote the support of a random vector $A$ and ${\cal S}_{A|b}$ the support of $A$ conditional on the value $b$ of a random vector $B$.  Moreover, to simplify, we assume that the upper bounds $\overline{\theta}(x)$, $\overline{a}(x)$, and $\overline{d}(x)$ are finite for every $x\in{\cal S}_X$.  Hence, all moments of $J$ exist and its moment-generating function is well-defined.
Otherwise, our identification arguments  hold  straightforwardly using  characteristic functions.

Hereafter, we show how   coverage choices combined with  sufficient variations in some exogenous characteristics 
nonparametrically  identify  $f(\theta,a|X)$  thereby offering  flexibility on the dependence between 
risk and risk aversion.\footnote{Using a log-normal joint distribution for $(\theta,a)$, Cohen and Einav (2007) find a counterintuitive positive correlation suggesting that the observed contracts are suboptimal, i.e.,  the insurer could increase his profit by increasing their low deductibles which are more compatible with a negative correlation.} 
To begin,  the damage distribution $H(\cdot|X)$ is identified  on its support
$(0,\overline{d}(X))$ using A2-(iv) 
since all the accidents and   damages are observed.  It remains to identify the joint distribution $F(\theta,a|X)$.

\bigskip\noindent
{\sc Identification of $F_{\theta|X}(\cdot|\cdot)$}

The first step identifies the   conditional distribution of risk $F_{\theta|X}(\cdot|\cdot)$ by exploiting the observed number of accidents/claims  $J$ and its nonparametric mixture. Specifically, the probability of $J$ conditional on the characteristics $X=x$ is
\begin{eqnarray*}
{\rm Pr}[J=j|x]=\int_{\underline{\theta}(x)}^{\overline{\theta}(x)} {\rm Pr}[J=j|\theta, x]dF_{\theta|X}(\theta|x) = \int_{\underline{\theta}(x)}^{\overline{\theta}(x)} e^{-\theta} \frac{\theta^j}{j!} dF_{\theta|X}(\theta|x), 
\end{eqnarray*} 
where the mixing distribution $F_{\theta|X}(\cdot|x)$ is left unspecified. 

 For  insurees with characteristics $x$, the moment-generating function
$M_{J|X}(\cdot|x)$ of the number of claims  is
\begin{eqnarray*}
M_{J|X}(t|x) &=& {\rm E}[e^{Jt}|X=x] = {\rm E}\left\{{\rm E}[e^{Jt}|\theta,X]|X=x\right\} \nonumber\\
&=& {\rm E}\left\{{\rm E}[e^{Jt}|\theta]|X=x \right\} 
=  {\rm E} \left\{e^{\theta(e^{t}-1)}|X=x \right\} \nonumber\\
&=& M_{\theta|X} (e^{t}-1 | x),  
\end{eqnarray*}
where the third and fourth equalities follow from A2-(iii) and the 
moment-generating function of the Poisson distribution with parameter $\theta$. 
In particular, this equation shows that  $M_{J|X}(\cdot|x)$ exists on $\Real$ because the moment-generating function $M_{\theta|X}(\cdot|x)$ of $\theta$ given $X=x$  exists on $\Real$ as $\theta$ has  compact support given $X=x$. 
Moreover, letting $u=e^t-1$ gives
\begin{eqnarray}
M_{\theta|X}(u|x)= M_{J|X}(\log(1+u)|x)= {\rm E}[(1+u)^J|X=x]
\end{eqnarray}
for all $u \in (-1,+\infty)$. Hence $M_{\theta|X}(\cdot|x)$ is identified in a neighborhood of $0$, thereby identifying  the density $f_{\theta|X}(\cdot|\cdot)$ of $\theta$ given $X$. See, e.g., Billingsley (1995, p.390).\footnote{Alternatively, because $M_{\theta|X}(\cdot|x)$ exists in a neighborhood of 0, then all the moments of $\theta$ given 
$X=x$ are identified by the $k$th derivatives $M^{(k)}_{\theta|X}(0|x)={\rm E}[\theta^k|X=x]$ for  $k=0,1,\ldots,\infty$. Since $\theta$ given $x$ has a bounded support, we are in the class of Hausdorff moment problems, which are always determinate, i.e., the distribution of $\theta$ given $x$ is uniquely determined by its moments.  For a comprehensive treatment of the moment problem, see Shohat and Tamarkin (1943).}  
This result exploits the fact that the Poisson distribution belongs to the class of additively closed distributions
whose nonparametric mixture is identified. See  Rao (1992) and  the online appendix.

\bigskip\noindent
{\sc Identification of $F_{a|\theta,X}[a(\theta,X)|\theta,X]$ }

The second step considers the probability that a $\theta$-insuree with characteristics $X$ chooses the coverage $(t_1(X),dd_1(X))$. 
We define a discrete variable $\chi$, which takes  values 1 and 2  depending on whether the insuree chooses the coverage $(t_1(X),dd_1(X))$
or  $(t_2(X),dd_2(X))$, i.e., whether his/her pair $(\theta,a)$ belongs to the regions ${\cal C}_1(X)$ or ${\cal C}_2(X)$ of individuals choosing contract 1 or 2, respectively given characteristics $X$. 
Thus, from Lemma 1-(ii), $\chi=1$ is also equivalent to $a\leq a(\theta,X)$ where  $a(\cdot,X)$ is the inverse of the frontier $\theta(\cdot,X)$ with
\begin{eqnarray} 
\theta(a,X)
   &=& \frac{t_2(X)-t_1(X)}{ \int_{dd_2(X)}^{dd_1(X)} e^{aD} (1-H(D|X)) dD}
 \end{eqnarray}
from (3).
Our identification strategy  exploits variations of this frontier in $X$.  In particular,  even if the deductible does not vary with $X$, the premium and/or the damage distribution likely depend  on some $X$.  

To ensure that the frontier (5) partitions $\Theta(X)\times{\cal A}(X)$ into the two nonempty sets ${\cal C}_1(X)$ and ${\cal C}_2(X)$, we assume that $\underline{\theta}(X) <\theta(\underline{a}(X),X)$ and $\theta(\overline{a}(X),X) < \overline{\theta}(X)$.  In particular,  ${\cal C}_1(X)$ includes the lowest type individual $(\underline{\theta}(X),\underline{a}(X))$ while ${\cal C}_2(X)$ includes the highest type individual $(\overline{\theta}(X),\overline{a}(X))$.    For $j=1,2$, let $\nu_j(X)$ denote the proportion of insurees with characteristics $X$ choosing the coverage $(t_j(X),dd_j(X))$.  Thus, $\nu_j(X)>0$ for $j=1,2$.  Such proportions are identified from the data.

The  probability ${\rm Pr}[\chi=1|\theta,X=x]$ that a $(\theta,x)$-individual chooses the lowest coverage contract $(t_1,dd_1)$ is
\begin{eqnarray}
F_{a|\theta,X}[a(\theta,x)|\theta,x] = \frac{f_{\theta|\chi,X}(\theta|1,x)\nu_1(x)}{ f_{\theta|X}(\theta|x)}
\end{eqnarray}
by Bayes' rule.
Since $f_{\theta|X}(\cdot|\cdot)$ is identified from the first step and $\nu_1(x)$ is identified from the data, it remains to identify 
$f_{\theta|\chi,X}(\cdot|1,x)$.
Applying the same argument as in Step 1, but  now conditioning on $\chi=1$ as well, we obtain 
\begin{eqnarray*}
M_{J|\chi,X}[t|1,x] &=& {\rm E}[e^{Jt}|\chi\!=\!1,X\!=\!x] = {\rm E}\{{\rm E}[e^{Jt}|\theta,a,X]|\chi\!=\!1,X\!=\!x\} \nonumber\\
&=&  M_{\theta|\chi,X}[e^t-1|1,x],   
\end{eqnarray*}
where the second equality follows from the equivalence between  conditioning on $(\theta,a,\chi,$ $X)$ and conditioning on $(\theta,a,X)$, while the third equality 
follows from A2-(iii) as before.
Thus, for every  $x \in {\cal S}_{X}$,  $f_{\theta|\chi,X}(\cdot|1,x)$ is identified by its moment generating function 
\begin{eqnarray*}
M_{\theta|\chi,X}(u|1,x)= M_{J|\chi,X}(\log(1+u)|1,x)
\end{eqnarray*}
for all $u \in (-1,+\infty)$.
Hence,  by (6), $F_{a|\theta,X}[a(\theta,x)|\theta,x]$ is identified for every $\theta \in (\underline{\theta}(x),$  $\overline{\theta}(x))$. That is, we
 identify the conditional distribution of $a$ given $\theta$ 
on the frontier $a(\theta,x)$  separating the two subsets ${\cal C}_1(x)$ and ${\cal C}_2(x)$ that partition   the set $\Theta(x) \times {\cal A}(x)$.

\bigskip\noindent
{\sc Identification of $F(\theta,a|X)$}

For  policy  counterfactuals
the analyst needs  to identify $F(\cdot,\cdot|x)$ on the whole support $\Theta(x) \times {\cal A}(x)$. This constitutes the third step of identification in which we make an exclusion restriction and a support assumption 
involving  some  characteristics $Z$  included in $X$  to achieve identification of the distribution $F_{a|\theta,X}(\cdot|\cdot,\cdot)$  on its support.

We  partition the vector of the insuree or car characteristics $X$ into $(X_0,Z)$.  

\medskip\noindent
{\bf Assumption A3:} {\em  We assume that $X$ satisfies the following\\
(i) $a \perp Z \big{|} (\theta,X_0)$\\
(ii) $\forall (a,\theta, x_0) \in {\cal S}_{a \theta X_0}$, there exists $z \in {\cal S}_{Z|\theta x_0}$ such that 
$a(\theta,x_0,z)=a$.}
 
\medskip\noindent
Assumption A3-(i) is an exclusion restriction. It requires  that  some variable    $Z$ is independent of  risk aversion conditionally on the other variables $X_0$ (and risk $\theta$).  
Assumption A3-(ii) is a full support assumption that requires  the frontier $a(\theta,X_0,Z)$ to vary sufficiently with $Z$. In particular, $Z$ needs to be continuous since $a$ is continuous.

In the case of automobile insurance, several  variables are potential candidates for $Z$.  For instance, controlling  insurees' characteristics such as age and others, Cohen and Einav (2007)   empirically find that the engine size and the years of license  are not related to  risk aversion.      Under the CARA specification, the car value, which acts as a proxy for wealth, could  also be a good candidate since CARA risk aversion is independent of wealth.  Nonetheless, these variables affect the frontier (5) through the  premia, the deductibles, and the distribution of damages.  Indeed, variations in the frontier arise from variations in premia but also through the difference in expected losses. For instance,  a large engine  or car value is less likely to lead to damage in the interval $[dd_2,dd_1]$. Thus, a large engine  or car value will give a larger value in the denominator of (5) than a low engine  or car  value. 

The combination of an exclusion restriction and a full support assumption is not new in the econometrics  literature. See, e.g., Matzkin (2003) and Imbens and Newey (2009). In empirical industrial organization, this includes Berry and Haile (2014) in the nonparametric identification of a demand and supply model for differentiated products. In auctions with selective entry,  Gentry and Li (2014) and Chen, Gentry, Li and Lu (2023) assume  the existence of a continuous entry cost shifter with full support  that affects entry but not the private value distribution. In an application to oil tract lease auctions, Kong (2018) finds that the amount of land offered for auction outside the area is an entry cost shifter satisfying such requirements.

Given  A3,  for any $(a,\theta,x_0) \in {\cal S}_{a \theta  X_0}$ we have 
\begin{eqnarray*}
F_{a|\theta,X_0}(a|\theta,x_0) = F_{a|\theta,X_0}[a(\theta,x_0,z)|\theta,x_0] = F_{a|\theta,X_0,Z}[a(\theta,x_0,z)|\theta,x_0,z],
\end{eqnarray*}
where the first equality uses A3-(ii) and the second equality uses A3-(i).
Note that $a(\cdot,\cdot,\cdot)$ is identified from  (5) since the premia and deductible are observed while the distribution of damage $H(\cdot|\cdot)$ is identified from claim data.
Identification of $F(\theta,a|x_0,z)$ follows from the identification of $F_{a|\theta,X}[a(\theta,x)|\theta,x]$ in Step 2 where $x=(x_0,z)$. This result is formally stated in the next proposition.

\medskip\noindent
{\bf Proposition 1:} {\em   Suppose two offered coverages  and  damages are observed for each insuree. Under A2--A3, the structure $[F(\cdot,\cdot|X),H(\cdot|X)]$ is identified.}

\medskip\noindent
Despite pooling, due to  multidimensional types and a finite number of coverages, Proposition 1 shows that the model primitives are identified by exploiting  the number of accidents and  sufficient variations in some exogenous variable $Z$ conditionally  independent of  risk aversion. In particular, our identification argument does not require optimality of the offered coverages. This argument  is novel  in the identification of models under incomplete information.

In the absence of the full support assumption A3-(ii), the previous argument shows that  we can still point identify $F_{\theta|X}(\cdot|\cdot)$ on ${\cal S}_{\theta X}$ as well as   the conditional distribution $F_{a|\theta, X}(a|\theta, x)= F_{a|\theta,X_0}(a|\theta, x_0)$ on  the range  of $a(\theta,x_0,\cdot)$ when $z$ varies, i.e. on
$\{ (a,\theta,x_0):  a=a(\theta,x_0,z),  z\in {\cal S}_{Z|\theta x_0 },  (\theta,x_0) \in {\cal S}_{\theta X_0}\}$. See also Section 4.  Moreover, 
assuming that this range is an interval 
$[a_*(\theta,x_0),a^*(\theta,x_0)]$, where $a_*(\theta,x_0)=\inf_{z \in {\cal S}_{Z|\theta x_0}} a(\theta,x_0,z)$ and 
$a^*(\theta,x_0)= \sup_{z \in {\cal S}_{Z|\theta x_0}} a(\theta,x_0,z)$, we can bound the conditional distribution $F(a|\theta,x_0)$  by  
\begin{eqnarray*}
0 \leq F_{a|\theta,X_0}(a|\theta,x_0) \leq 
F_{a|\theta,X_0}(a_*(\theta,x_0)|\theta,x_0) \\
{\rm and} \quad   F_{a|\theta,X_0}(a^*(\theta,x_0)|\theta,x_0) \leq 
F_{a|\theta,X_0}(a|\theta,x_0) \leq 1
\end{eqnarray*}
for $\underline{a}(x_0) \leq a \leq a_*(\theta,x_0)$
and $a^*(\theta,x_0) \leq a \leq \overline{a}(x_0)$, respectively.  These bounds are sharp as there is no information on
$[\underline{a}(x_0),a_*(\theta,x_0))$ and 
 $(a^*(\theta,x_0),  \overline{a}(x_0)]$.\footnote{Formally, let 
$\tilde{F}_{a|\theta,X_0}(\cdot|\theta,x_0)$ be another distribution that differs from $F_{a|\theta,X_0}(\cdot|\theta,x_0)$  only on $(\underline{a}(\theta,x_0), a_*(\theta,x_0))\cup (a^*(\theta,x_0),\overline{a}(\theta,x_0))$. These two distributions lead to the same  probability ${\rm Pr}(\chi=1|\theta,X=x_0,Z=z)= {\rm Pr}(a \leq a(\theta,x_0,z)|\theta, X=x_0)$ since the frontier $a(\cdot,x_0,z)$ does not depend on this conditional distribution by (5).}
It should  also be noted that having a larger number of coverages $C >2$ can only improve the identification results as  a larger number of frontiers of the form (5)
is more likely to cover the whole support $\Theta(x) \times {\cal A}(x_0)$ when 
$Z$ varies as discussed  next.

\bigskip\noindent
{\sc Beyond Two Coverages}

We now consider more than two offered coverages.
Let $(t_c,dd_c)$, $c=1,\ldots,C\geq 2$, be $C$ offered contracts. We omit the insuree/car characteristics to simplify the notations.  As before, we  require  that no observed coverage dominates the others:
\begin{eqnarray}
0 <t_1<\ldots<t_C \ {\rm and} \  
\overline{d}>dd_1>\ldots>dd_C\geq 0 .
\end{eqnarray}
We refer to (7) as the revealed preference (RP) condition, since otherwise some  contracts will be irrelevant. It extends  the condition that we have for $C=2$. Moreover, this condition is easily verifiable in the data.
Let $\theta_{c,c+1}(a)$ define the frontier or indifference locus between coverages $(t_c,dd_c)$ and $(t_{c+1},dd_{c+1})$.  It is given by an equation similar to (3).   By Lemma 1-(i),  each frontier is decreasing and $(\theta,a)$-individuals below (resp. above) the curve $\theta_{c,c+1}(\cdot)$ prefers coverage $(t_c,dd_c)$ over coverage $(t_{c+1},dd_{c+1})$ (resp. $(t_{c+1},dd_{c+1})$ over $(t_{c},dd_{c})$).

The next lemma ensures that the $C-1$ frontiers $\theta_{c,c+1}(\cdot)$ do not cross and lie on top of each other as $c$ increases from 1 to $C-1$.

\medskip\noindent
{\bf Lemma 2:} {\em Let A1 hold and the coverages $(t_c,dd_c)$,  $c=1,\ldots, C\geq 2$ satisfy the RP condition (7).  The frontiers $\theta_{c,c+1}(\cdot)$ between coverages $(t_c,dd_c)$ and $(t_{c+1},$ $dd_{c+1})$ for $c=1,\dots,C-1$ satisfy $\theta_{1,2}(\cdot)<\ldots<\theta_{C-1,C}(\cdot)$ 
on $[\underline{a},\overline{a}]$ if and only if 
\begin{eqnarray}
\frac{t_{c+2}-t_{c+1}}{t_{c+1}-t_c} > 
\frac{ \int_{dd_{c+2}}^{dd_{c+1}} e^{\underline{a}D}[1-H(D)]\ dD }
{\int_{dd_{c+1}}^{dd_{c}} e^{\underline{a}D}[1-H(D)]\ dD} 
\end{eqnarray}
for $c=1,\ldots,C-2$. 
}

\medskip\noindent
Condition (8) depends on the terms of the offered contracts as well as on the damage distribution.  In contrast, it does not depend on the distribution of risk and risk aversion except through the lower bound $\underline{a}$ of risk aversion. 

Interestingly, if $\underline{a}$ approaches zero,  condition (8) becomes
\begin{eqnarray*}
\frac{t_{c+2}-t_{c+1}}{t_{c+1}-t_c} > 
\frac{ \int_{dd_{c+2}}^{dd_{c+1}} [1-H(D)]\ dD }
{\int_{dd_{c+1}}^{dd_{c}}[1-H(D)]\ dD} 
\end{eqnarray*}
for $c=1,\ldots,C-2$.  Applying the Mean Value Theorem gives
\begin{eqnarray*}
\frac{t_{c+2}-t_{c+1}}{t_{c+1}-t_c} > \kappa_{c+1}
\frac{dd_{c+1}-dd_{c+2}}{dd_{c}-dd_{c+1}},
\end{eqnarray*} 
where $\kappa_{c+1}=[1-H(D_{c+1}^*)]/[1-H(D_{c}^*)]>0$ with $D_{c+1}^* \in (dd_{c+2},dd_{c+1})$ and $D_{c}^* \in (dd_{c+1},dd_{c})$.  
In particular, because $1-H(D)$ is decreasing in $D$,  we have $\kappa_{c+1} >1$.  Thus the increments in premia should increase proportionally more than the decrements in deductibles. This relates to a well-known property of reverse nonlinear pricing as noted by Stiglitz (1977). 
The next corollary formalizes this result.

\medskip\noindent
{\bf Corollary:} {\em  Let A1 hold and the coverages $(t_c,dd_c), c=1\ldots,C \geq 2$ satisfy the RP condition (7).  When $\underline{a}$ approaches zero,  a necessary and sufficient condition for (8) is 
\begin{eqnarray}
\frac{t_{c+2} - t _{c+1}}{|dd_{c+2}-dd_{c+1}|}   > \kappa_{c+1}
\frac{t_{c+1} - t _{c}}{|dd_{c+1}-dd_{c}|},
\end{eqnarray}
for some $\kappa_{c+1}>1$ and $c=1,\ldots,C$.}

\medskip\noindent
 This corollary  says that the observed coverages $(t_c,dd_c), c=1,\ldots,C$ should lie on a convex curve in the $(t,dd)$-space. 
This convexity  is easily verifiable in the data. It should be noted that  such a theoretical property is obtained here despite  non optimal   contracts and bidimensional incomplete information.  See also Luo, Perrigne and Vuong (2017, 2018).   

When either  (8) or (9) holds, any individual whose type $(\theta,a)$ lies between the frontiers $\theta_{c,c+1}(\cdot)$ and $\theta_{c+1,c+2}(\cdot)$ chooses the  coverage $(t_{c+1},dd_{c+1})$, for $c=1,\ldots,C-2$.  Indeed, from Lemma 1,  this individual prefers $(t_{c+1},dd_{c+1})$ to $(t_{c+2},dd_{c+2})$,   which is preferred to $(t_{c+3},dd_{c+3})$, etc.  Similarly, this individual prefers $(t_{c+1},dd_{c+1})$ to $(t_{c},dd_{c})$,  which is preferred to $(t_{c-1},dd_{c-1})$, etc.  Thus, a $(\theta,a)$-individual above the $\theta_{C-1,C}(\cdot)$-frontier chooses the highest coverage (i.e., the lowest deductible) $(t_C,dd_C)$.   Figure 1  illustrates the choice among the three contracts $(t_1,dd_1)=(600,1000)$, $(t_2,dd_2)=(850,500)$ and $(t_3,dd_3)=(1000, 250)$, which satisfy  condition (9).  In contrast to the case with two coverages,  insurees who are on the right  of the frontier  2 versus 3  now choose  $(t_3,dd_3)$. 

Our previous identification results  extend to more than two contracts. Specifically, under A2, the first step that identifies the marginal distribution $F_{\theta|X}(\cdot|\cdot)$ from the observed number of accidents remains the same as before.
The second step identifies the conditional distribution $F_{a|\theta,X}(\cdot|\cdot,\cdot)$ at the $C-1$ frontiers between coverages  $c$ and $c+1$ for $c=1,\ldots,C-1$ upon introducing the choice variable $\chi$ taking values $1,\ldots, C$
and the corresponding  proportions of individuals choosing coverage $c=1,\ldots,C$.
Hence, under the exclusion and support assumption A3, the distribution $F(\theta,a|X)$ is identified. As a matter of fact, A3-(ii) is stronger than necessary as it suffices  that the combined variations of the $C-1$ frontiers cover the $\Theta(X) \times {\cal A}(X_0)$ space. 
Moreover, if this sufficient condition is not satisfied, 
$F_{a|\theta,X}(a|\theta,x)$ is identified on  a larger set through the variations of the $C-1$ frontiers.
Thus, having more coverages helps identify the joint distribution of types $(\theta,a)$. 

In most of the empirical literature on insurance, such as in Israel (2005), Cohen and Einav  (2007), and
Barseghyan et al.  (2013), data are collected from a single company. In this case, our results  immediately apply.  If data combine insurees from different firms,  our approach requires that the observed coverages satisfy the revealed preference condition (7) arising from individuals' choices. This condition might not hold  if 
switching costs are present, preventing individuals from changing to 
their  preferred coverages.  However, when insurance contracts are differentiated in other dimensions, such as vehicle replacement, uninsured motorists, or roadside assistance, the analysis can be performed by conditioning on these add-ons. It then suffices to mix the recovered conditional distributions with the proportions of individuals choosing these add-ons to   obtain the joint distribution  $F(\cdot,\cdot)$ of risk and risk aversion.

\section{Estimation Method and Monte Carlo Study}

This section presents a computationally friendly three-step nonparametric  procedure for estimating the joint density $f(\theta,a|X)$ of risk $\theta$ and risk aversion $a$  given $X$ with support $[\underline{\theta},\overline{\theta}] \times [\underline{a},\overline{a}]= [0,1] \times [0,\overline{a}]$.\footnote{The interval $[0,1]$ is a normalization, whereas $\overline{a}$ is chosen by the analyst.} To simplify, we omit the covariates $X_0$ in $X=(X_0,Z)$, which can be entertained by conditioning our estimation procedure   on $X_0$ through a smoothing method such as  kernel estimation. Our procedure follows our identification argument as the latter is constructive. Hereafter, we consider two coverages. Let $(\chi_i,J_i, D_{1i}, \ldots,D_{J_i i},Z_i), i=1,\ldots,N$ be the available data where $\chi_i=c$ if individual chooses coverage $c=1,2$.  The estimation method is implemented in a Monte Carlo study.

\subsection{A Three-Step Estimation Procedure}

Our estimation procedure consists in three steps:

\noindent
Step 1:  Estimate  $f_{\theta|Z}(\cdot|\cdot)$ by  constrained Generalized Method of Moments (GMM) 

$\hspace{0.7cm}$
and kernel smoothing,

\noindent
Step 2: Estimate $f_{\theta|\chi,Z}(\cdot|1,\cdot)$  by  adapting Step 1 and conditioning on $\chi=1$,

\noindent
Step 3: Estimate $f_{a|\theta}(\cdot|\theta)$ by  plugging-in  estimates of $\partial \theta(a,z)/\partial a$, $\partial \theta(a,z)/\partial z$ and 

$\hspace{0.7cm}$
$\partial {\rm Pr}(\chi=1|\theta(a,z),z)/\partial z$.

\noindent
Hereafter, we present these steps in details.

\bigskip\noindent
{\sc Estimation of $f_{\theta|Z}(\cdot|\cdot)$}

Following the identification argument, 
the first step estimates the density $f_{\theta|Z}(\theta|z)$ of risk $\theta$ given $Z$.  To fix ideas,  we  assume  that this density does not depend on  $Z$.  The problem  then reduces to estimating the mixing distribution in  a Poisson mixture because  the observed numbers of accidents  $J_i, i=1,\ldots,N$  are i.i.d. drawn from  ${\rm Pr}(J=j)= \int_{\underline{\theta}}^{\overline{\theta}} [e^{-\theta} \theta^j/j!] f_\theta(\theta) d\theta$. 
The statistical literature views the estimation of the Poisson mixing density $f_\theta(\cdot)$ as a  Hausdorff moment problem using the empirical  (raw) moments $\hat{\mu}_m, m=1,\ldots,M\geq 1$
\begin{eqnarray*}
\hat{\mu}_m = \frac{1}{N} \sum_{i=1}^N J_i (J_i-1) \ldots
(J_i-m+1) \quad {\rm for} \ m\geq 1
\end{eqnarray*}
from,  e.g., Hengartner (1997).
 Specifically,  following Talenti (1987),  the estimator for $f_{\theta}(\theta)$  is
\begin{eqnarray*}
\hat{f}_\theta(\theta;\hat{\lambda})= 1 +\hat{\lambda}_1 L_1(\theta) + \ldots + \hat{\lambda}_M L_M(\theta),
\end{eqnarray*}
where $L_m(\theta)= c_{m0} + c_{m1} \theta + \ldots+ c_{mm} \theta^m$ is the  shifted Legendre polynomial of degree $m$ on $[0,1]$. The first coefficient is equal to one to satisfy $\int_{0}^1 \hat{f}_\theta(\theta;\lambda) d\theta=1$, and 
$\hat{\lambda}_m= c_{m0}+ c_{m1} \hat{\mu}_1 + \ldots + c_{mm} \hat{\mu}_m$ for $m=1,\ldots,M$. This follows by solving the $M$ equations
\begin{eqnarray}
\int_0^1 \theta^m \hat{f}_\theta(\theta;\lambda) d\theta = \hat{\mu}_m, m=1, \ldots, M.
\end{eqnarray}
 Hengartner (1997) shows that the optimal  convergence  rate is attained when $M= \log N/$ $\log(\log N)$. This rate is  slow relative to $\sqrt{N}$, reflecting  the difficulty of estimating $f_\theta(\cdot)$.\footnote{An alternative estimator consists in inverting the empirical characteristic function of $\theta$ by Fourier inversion. See Aryal, Perrigne and Vuong (2019). The deconvolution estimator implicitly  requires that moments are well estimated. In the case of automobile insurance, the number of accidents tends to be small rendering estimation of moments above four or five very imprecise.} 

To improve the finite sample properties,  we impose  that the  resulting estimated density is nonnegative through a constrained GMM estimator.\footnote{We are grateful to Matheus Silva for proposing this method. See Silva (2024).} Specifically, $\hat{\lambda}=(\hat{\lambda}_1,\ldots,\hat{\lambda}_M)$ is obtained as
\begin{eqnarray*}
\hat{\lambda} = {\rm argmin}_{\lambda=(\lambda_1,\ldots,\lambda_M)}
[\hat{\mu}- \mu(\lambda)]' V^{-1}[\hat{\mu}- \mu(\lambda)],
\end{eqnarray*}
subject to $\hat{f}_\theta(\theta;\lambda)  \geq 0$, where 
$\hat{\mu}=(\hat{\mu}_1,\ldots,\hat{\mu}_M)$ and 
$\mu(\lambda)= (\mu_1(\lambda),\ldots,\mu_M(\lambda))$ with
$\mu_m(\lambda)= \int_0^1 \theta^m  \hat{f}_\theta(\theta;\lambda) d\theta$. The weighting matrix is  $V={\rm diag}(\widehat{{\rm Var}}(\hat{\mu}_m))$, where
\begin{eqnarray*}
\widehat{{\rm Var}} (\hat{\mu}_m) = \frac{1}{N^2} \sum_{i=1}^N 
[J_i (J_i-1)\ldots (J_i-m+1)]^2 - \frac{1}{N} \hat{\mu}^2_m.
\end{eqnarray*}
When the marginal density of $\theta$ depends on $Z$, we  adapt the estimator by conditioning on $Z=z$ upon considering 
the empirical conditional moment $\hat{\mu}_m(z)$ obtained from a nonparametric regression of $J_i(J_i-1)\ldots(J_i-m+1)$ on $Z_i$ since ${\rm E}[\theta^m|Z] = {\rm E}[J(J-1)\ldots(J-m+1)|Z]$. This gives $\hat{f}_{\theta|Z}(\theta|z)$ upon applying the above estimator for each $z$ value.

\medskip\noindent
{\sc Estimation of $f_{\theta|\chi,Z}(\cdot|1,\cdot)$}

The third step  requires an estimator of   the conditional choice probability 
${\rm Pr}[\chi=1|\theta,z]$ that an individual  with risk $\theta$ and covariates $Z=z$ chooses coverage 1.  This probability is given by (6), whose right-hand side involves $f_{\theta|\chi,Z}(\theta|1,z)$,  $\nu_1(z)$ and $f_{\theta|Z}(\theta|z)$. 
The term $\nu_1(z)$ is the probability  of choosing coverage 1 given $Z=z$, and can be estimated by a nonparametric regression 
of $\chi_i$ on $Z_i$.  The  conditional density  $f_{\theta|Z}(\theta|z)$ is estimated by $\hat{f}_{\theta|Z}(\theta|z)$ obtained in the first step. It remains to estimate $f_{\theta|\chi,Z}(\cdot|1,z)$. 
A natural method  would apply  the first step estimator on the subsample of individuals choosing coverage 1.
However, this method  suffers from a possible  irregularity  at $\theta(\overline{a},z)$ when the latter belongs to $(0,1)$ as displayed  in Figure 1 or
Cohen and Einav (2007, Figure 2).\footnote{A formal proof is available upon request from the authors.}  

To address this difficulty, we note that by Bayes rule we have $f_{\theta|\chi,Z}(\theta|1,z)= f_{\theta|Z}(\theta|z)/\nu_1(z)$
when $\theta \in [0, \theta(\overline{a},z)]$ because all such  $\theta$-individuals always choose coverage 1. Thus, from above $f_{\theta|\chi,Z}(\cdot|1,z)$ is readily estimated on $[0, \theta(\overline{a},z)]$.
It remains to estimate this density on $[\theta(\overline{a},z), \min \{\theta(0,z),1 \}]$. We have
\begin{eqnarray}
f_{\theta|\chi,Z}(\theta|1,z) = 
\left[1 - \frac{F_{\theta|Z}(\theta(\overline{a},z)|z)}{\nu_1(z)}  \right] g(\theta|1,z) 
\end{eqnarray}
when $\theta \in [\theta(\overline{a},z), \min \{\theta(0,z),1 \}]$,
where $g(\cdot|1,z)$ is the conditional density of $\theta$  given  $\{\theta>\theta(\overline{a},z),\chi=1,Z=z\}$.  Thus the problem reduces to  estimating  the density $g(\cdot|1,z)$.  The support boundaries 
$\theta(\overline{a},z)$ and $\theta(0,z)$ are estimated by letting $a=\overline{a}$ and $a=0$, respectively, in the estimated frontier 
\begin{eqnarray}
\hat{\theta}(a,z) = \frac{t_2(z) - t_1(z)}{\int_{dd_2(z)}^{dd_1(z)} e^{aD} [1- \hat{H}(D|z)] dD},
\end{eqnarray}
where $\hat{H}(D|Z)$ is  a nonparametric estimator of the   damage distribution.

To estimate $g(\theta|1,z)$, we apply the first-step estimator   accounting for its support $[\theta(\overline{a},z),\theta(0,z)]$.\footnote{We assume that $\theta(0,z) \leq 1$ as in Figure 1 and in Cohen and Einav (2007, Figure 2).} Specifically, let
\begin{eqnarray*}
g(\theta|1,z)&=& \frac{1}{ \theta(0,z)-\theta(\overline{a},z)} \\
 && \times \left[ 1 + \beta_{1z} L_1\left(\frac{\theta- \theta(\overline{a},z)}{\theta(0,z)-\theta(\overline{a},z)}  \right) + \ldots+
\beta_{Mz}   L_M\left(\frac{\theta- \theta(\overline{a},z)}{\theta(0,z)-\theta(\overline{a},z)}  \right)   \right],
\end{eqnarray*}
where the coefficients $(\beta_{1z},\ldots,\beta_{Mz})$ depend on $z$. By (11),
the $m$th moment of $g(\theta|1,z)$ is obtained from  the $m$th moment 
of $f_{\theta|\chi,Z}(\theta|1,z)$ using
\begin{eqnarray*}
 {\rm E} (\theta^m|\chi=1,z) =\int_0^{\theta(\overline{a},z)} \theta^m \frac{f_{\theta|Z}(\theta|z)}{\nu_1(z)} d\theta + \left[1- \frac{F_{\theta|Z}(\theta(\overline{a},z)|z)}{\nu_1(z)} \right]
\int_{\theta(\overline{a},z)}^{\theta(0,z)} \theta^m g(\theta|1,z) d\theta,
\end{eqnarray*}
for $m=1,\ldots,M$.
The estimated parameters  $(\hat{\beta}_{1z},\ldots,\hat{\beta}_{Mz})$ are obtained by GMM subject to the constraints  
\begin{eqnarray*}
&& (i) \quad 0 \leq \left[ 1 - \frac{F_{\theta|Z}(\theta(\overline{a},z)|z)}{\nu_1(z)}\right]   g(\theta|1,z)  \leq \frac{f_{\theta|Z}(\theta|z)}{\nu_1(z)},\\
&& (ii)\ \  \left[ 1 - \frac{F_{\theta|Z}(\theta(\overline{a},z)|z)}{\nu_1(z)}\right] g(\theta(\overline{a},z)|1,z) = \frac{f_{\theta|Z}(\theta(\overline{a},z)|z)}{\nu_1(z)},
\end{eqnarray*}
upon replacing  $\theta(0,z), \theta(\overline{a},z),$  $f_{\theta|Z}(\theta|z), 
F_{\theta|Z}(\theta|z), \nu_1(z)$ and ${\rm E} (\theta^m|\chi=1,z)$ by their estimated counterparts.\footnote{The first constraint follows from $f_{\theta|Z}(\theta|z)= f_{\theta|\chi,Z}(\theta|1,z) \nu_1(z) + 
f_{\theta|\chi,Z}(\theta|2,z) \nu_2(z) \geq f_{\theta|\chi,Z}(\theta|1,z) \nu_1(z)$ and (11).  The second constraint  imposes the continuity of 
$f_{\theta|\chi,Z}(\cdot|1,z)$ at $\theta(\overline{a},z)$. To improve finite sample properties, we also impose  $g(\theta(0,z)|1,z)=0$  in Section 4.2.}  In particular, $\hat{{\rm E}} (\theta^m|\chi=1,z)$ is obtained from a nonparametric regression of $J_i(J_i-1)\ldots (J_i-m+1)$ on $Z_i$ using the subsample of individuals who choose coverage 1.  This gives the estimator $\hat{g}(\theta|1,z)$ and hence the estimator $\hat{f}_{\theta|\chi,z}(\theta|1,z)$ using (11). 
The estimator of ${\rm Pr}[\chi=1|\theta,z]$ is  $\widehat{{\rm Pr}}[\chi=1|\theta,z]=\hat{f}_{\theta|\chi,z}(\theta|1,z)\hat{\nu}_1(z)/ \hat{f}_{\theta|Z}(\theta|z)$.

\medskip\noindent
{\sc Estimation of $f_{a|\theta}(\cdot|\cdot)$}

The third step estimates the density $f_{a|\theta}(a|\theta)$ of risk aversion $a$, conditional on risk $\theta$  under the exclusion restriction in Assumption A3-(i). Our  argument of Section 3 identifies 
the distribution $F_{a|\theta}(\cdot|\theta)$ for those values of $a$ for which $a=a(\theta,z)$ for some $z$. This is inconvenient for estimating 
its density $f_{a|\theta}(\cdot|\theta)$. We exploit instead  the identity $F_{a|\theta}[a(\theta,z)|\theta]=\Pr[\chi=1|\theta,z]$.  Differentiate it with respect to $z$ and using $\partial a(\theta,z)/\partial z=- \{\partial \theta[a(\theta,z),z]/\partial z\}/\{\partial \theta[a(\theta,z),z]/\partial a\}$ give
\begin{eqnarray}
f_{a|\theta}[a(\theta,z)|\theta] = - \frac{\partial \theta[a(\theta,z),z]\partial a}{\partial \theta[a(\theta,z),z]/\partial z}
\times 
\frac{\partial {\Pr}[\chi=1|\theta,z]}{\partial z},
\end{eqnarray}
where $a(\cdot,z)$ is the inverse of $\theta(\cdot,z)$, and $\dim Z =1$ to simplify.
 In particular, we identify $f_{a|\theta}(\cdot|\theta)$ on the range of $a(\theta,z)$ when $z$ varies. Under the full support assumption A3-(ii), this range  is $[0,\overline{a}]$. 
To estimate $f_{a|\theta}(\cdot|\theta)$ at a value $a$ in the range of $a(\theta,z)$, we use numerical derivatives for $\partial \widehat{{\rm Pr}}[\chi=1|\theta,z]/\partial z$ and $\partial \hat{\theta}(a,z)/\partial z$, while $\partial \hat{\theta}(a,z)/\partial a$ is readily available from (5), where $H(D|z)$ is replaced by its estimate $\hat{H}(D|z)$.\footnote{The resulting  estimator  may not be positive. One can take its absolute value.
When the range is the full support, normalizing it by its integral over $[0,1]$  provides an estimator of $f_{a|\theta}(\cdot|\theta)$ satisfying the properties of a density.}  
With  $C>2$ contracts, there are $C-1$ frontiers $\theta_{c,c+1}(a,z)$, $c=1,\ldots,C-1$. Thus  $f_{a|\theta}(\cdot|\theta)$ can be estimated on a larger range of values of $a$ when $z$ varies.

\subsection{A Monte Carlo Study}

This section  implements the above estimator on simulated data. 

\medskip\noindent
{\sc Data-Generating Process}

We consider a  Monte Carlo setup  that    captures some basic features of automobile insurance data.  Risk $\theta$ and risk aversion $a$ are marginally distributed as Beta(2,3) on $[0,1]$  and $10^{-3}$Beta(1,3) on $[0,10^{-3}]$, respectively.  The range of values is similar to those found by Cohen and Einav (2007, Figure 1).  In agreement with the intuition that risk aversion  is associated with a tendency to take greater precautions, we allow for a negative association between risk and risk aversion through a Gaussian copula with correlation  $\rho=-0.5$. See Finkelstein and McGarry (2006), whose empirical results support this intuition.
Damages are  exponentially  and independently distributed  with a mean of 5,000, whereas the number $J$ of accidents is distributed as Poisson with parameter $\theta$. 
We present a simplified version of Assumption A3-(i) with an exogenous variable $Z$ that is uniformly distributed on $[100,200]$ and independently of $(\theta,a)$.
We consider two coverages with fixed deductibles at 1,000 and 500 with premia $3.25 Z$ and 700 for coverages 1 and 2, respectively.  Having fixed  deductibles is standard among insurance companies. Though $Z$ is independent of $(\theta,a,J,D)$, it enters in the insuree's contract choice  as he/she chooses coverage 1 if  his/her risk $\theta_i$ is below the frontier $\theta(a_i,Z_i)$, where  the frontier $\theta(a,Z)$ is given by (12).  This frontier varies in $Z$ through the numerator, which is sufficient for  identification and estimation. 

We draw a sample of 100,000 triplets $(\theta_i,a_i, Z_i)$ from  $F(\theta,a)$ and  $U(100,200)$.
The value of $Z_i$ determines the pair of offered coverages and the frontier (12).
Individual $i$ chooses coverage $(1,000, 3.25Z_i)$, i.e. $\chi_i=1$ if his/her risk $\theta_i \leq \theta(a_i,Z_i)$. Given $\theta_i$, a number  $J_i$ of accidents  is drawn from  a Poisson with mean $\theta_i$. Damages $D_{1i},\ldots,D_{J_ii}$ are drawn from $H(\cdot)$. 
Figure 2 displays  the observations $(\theta_i,a_i)$ for one simulated sample. The frontiers  $\theta(a,z)$  when $z$ varies from 110 (right curve)  to 190 (left curve) provide the locuses of points $(\theta,a)$ for which the  $z$-individuals   are indifferent between the two  coverages.
If a $z$-individual  has a $(\theta,a)$ pair under $\theta(\cdot,z)$, then he/she  chooses  coverage 1; otherwise, he/she  chooses the second coverage offering a better protection.  Figure 3 displays the histogram of the number of accidents. A large majority of individuals have no accident, and the proportions decline sharply to reach values close to zero for $J \geq 4$, in agreement with Cohen and Einav (2007, Table 2B).  Using this random sample, we perform the estimation procedure detailed above. We repeat this exercise 100 times. 

\medskip\noindent
{\sc Monte Carlo Results}

Figure 4 shows  the  estimated marginal density of the  expected number  $\theta$ of accidents. It displays the true density as well as the 90\% confidence interval.  The true curve is within the corresponding bounds, which are remarkably narrow. The  constrained GMM estimator is implemented using $M=4$ moments  which is the integer part of $\log N/(\log \log N)$.  To save space,  we do not display the results of Step 2 because it is an intermediary step where we use the estimator from (11) since the frontiers $\theta(a,z)$ have an  irregularity at $\theta(\overline{a},z) \in (0,1)$.   We apply the constrained GMM estimator  for the density $g(\theta|1,z)$ with $M=4$ moments  on its  support 
$[\theta(\overline{a},z),\theta(0,z)]$. This step also requires estimates of the probability $\nu_1(z)$ of choosing coverage 1  as well the  kernel regression  of $J_i(J_i-1)\ldots (J_i-m+1)$ on $Z_i$ with $m=1,\ldots, 4$. Kernel estimators are performed using rule-of-thumb bandwidths.

Figure 5 displays the density $\hat{f}_{a|\theta}(\cdot|0.4)$ conditional 
on $\theta=0.4$ since $\theta$  is distributed as $B(2,3)$ with mean 0.4.  
This density estimator is obtained from (13).   Figure 5 also provides the 90\% confidence interval,  which is relatively narrow and contains the true conditional density. It is worth noting that  the range of $a(0.4,z)$ is  $[0,10^{-3}]$ when $z$ varies. 
In contrast, Figure 6  displays  $\hat{f}_{a|\theta}(\cdot|0.6)$ conditional on $\theta=0.6$. We observe that the range of $a(0.6,z)$ is $[0,0.44 \times 10^{-3}]$. This finding  illustrates that the support assumption partially holds as  the variation in $z$ is not sufficient to estimate this conditional density at $\theta=0.6$ on its full support $[0,10^{-3}]$. As discussed previously, this issue could be alleviated when observing more than two insurance options.  Nonetheless, the 90\% confidence interval on the identified range contains the true density except at the  leftmost boundary.   Boundary effects are typical in nonparametric estimation and can be corrected.

\section{Conclusion}

Our paper addresses the identification and estimation of  insurance models  where insurees have private information about  their risk and risk aversion.  Our model also includes random damages and  the possibility of multiple accidents. Despite  bunching due to  multidimensional types and  a  finite number of offered coverages, we  identify the model primitives by exploiting the observed number of claims.  We also develop a  nonparametric estimation procedure  that is computationally friendly.
Our  results apply to any form of competition and do not rely on the optimality of offered coverages in contrast to the previous literature on the identification and estimation of models with private information. Thus, their optimality could be tested upon the specification of an appropriate model of market competition. 
Beyond optimality, several counterfactuals can be performed. For instance,   we can assess the gain/loss for both parties  from  (i)  reducing the range of   insurees' characteristics that the insurer can use to discriminate insurees such as gender,  age, or location, (ii) increasing the number of existing coverages and/or changing their terms, 
and (iii) implementing other coverages than premium/deductible with (say) a proportional deductible.

In terms of future lines of research, first,   our results extend to a broad range of insurance data, such as in health, provided the analyst observes repeated outcomes, e.g. insurees' claims.   In particular, we may want to extend our identification results  to allow for some form of moral hazard. Second, in the case of automobile insurance, we could endogenize
the car choice given insuree's risk and risk aversion. This extension would lead to  a model explaining the car choice, the coverage choice, the number of accidents, and the damages.  
Third, several existing data sets  on automobile and home insurance used by Israel (2005),  Cohen and Einav (2007), Sydnor (2010), and Barseghyan, Molinari, O'Donoghue and Teitelbaum (2013) could be analyzed using our empirical framework.

\newpage

\small

\bigskip
\begin{center}
{\bf Appendix}
\end{center}

\setcounter{equation}{0}
\renewcommand{\theequation}{A.\arabic{equation}}

\bigskip\noindent
The appendix contains the extension to health insurance as well as the proofs of Lemmas 1 and 2.

\bigskip\noindent
{\bf  The Case of Health Insurance:}
Up to some variations, health insurance  involves a premium $t$ as well as a per period deductible $dd$ and a copayment $\gamma$ per (say) medical visit. In particular, the deductible is not per visit, while the copayment arises on the first visit after the deductible is met. 
In this case,  when buying a contract $(t,dd,\gamma)$, the $(\theta,a)$-patient has an expected utility
\begin{eqnarray*}
V(t,dd,\gamma;\theta,a,w) = - e^{-a(w-t)} {\rm E}[e^{-a Y(dd,\gamma)}|\theta],
\end{eqnarray*}
where  the insuree's expense beyond the premium (commonly referred as out-of-pocket) is  $Y(dd,\gamma)= (D_1+\ldots+D_J) \Unit(D_1+\ldots+D_J \leq dd) + (dd + (J-J^{\dagger}) \gamma)\Unit(D_1+\ldots+D_J > dd)$ with $J$  the number of visits and  $J^{\dagger}$ 
the number of visits at which the deductible is met, i.e.,
$J^{\dagger} = {\rm argmin}_{j=1,\ldots,J} D_1+\ldots+D_j > dd$.
The expectation is with respect to the total expense $D_1+\ldots+D_J$ and the number $J$ of visits which depends on $\theta$.
In the case of health coverage,
the per visit expenses $D_j, j=1,\ldots,J$ may  be viewed as independent conditional on  the patient's health conditions such as cancer, diabetes, which are observed by the analyst. Similarly, conditioning on the patient's health conditions alleviates the possible dependence between $D_j$ and the expected number of medical visits $\theta$. See Assumption A2 in the text  for the introduction of insurees' characteristics.
Letting $m(dd,\gamma;\theta,a)={\rm E}[e^{-a Y(dd,\gamma)}|
\theta]$, the certainty equivalent becomes
\begin{eqnarray*}
CE(t,dd,\gamma;\theta,a,w)= w-t - \frac{\log m(dd,\gamma;\theta,a)}{a},
\end{eqnarray*}
which is similar to (2).

\bigskip\noindent
{\bf Proof of Lemma 1:}  The first part of (i) is immediate from (2) since $\phi_a(dd)=1$ when $dd=0$. 
When $dd>0$, the derivative of (2) with respect to $\theta$ is $-(\phi_a-1)/a$. Since $\phi_a >1$, $CE(t,dd;\theta,a,w)$ is decreasing in $\theta$.
For the derivative of (2) with respect to $a$, we note that 
\begin{eqnarray}
\phi_a(dd)=\int_{0}^{dd} e^{aD}dH(D)+e^{add}[1-H(dd)] = 1 + a\int_0^{dd}e^{aD}[1-H(D)] dD
\end{eqnarray}
by integration by parts.  Thus (2) gives 
$$
CE(t,dd;\theta,a,w)=w-t-\theta\int_0^{dd}e^{aD}[1-H(D)] dD.
$$
Hence, $CE(t,dd;\theta,a,w)$ is decreasing in $a$.

We now prove (ii).  The frontier between ${\cal C}_1$ and ${\cal C}_2$ is defined as the locus of $(\theta,a)$-insurees who are indifferent between the two coverages, i.e., for whom $CE(t_1,dd_1;\theta,a,w)=CE(t_2,dd_2;\theta,a,w)$. Using (2) this gives 
\begin{eqnarray*}
t_1+\frac{\theta[\phi_a(dd_1)-1]}{a} = t_2+\frac{\theta[\phi_a(dd_2)-1]}{a} .
\end{eqnarray*}
Solving for $\theta$ as a function of $a$ gives (3) upon using (A.1).  Moreover,
from (3) it is easy to see that $\theta(a)$ decreases in $a$.  The last part of (ii) also follows as $CE(t_1,dd_1;\theta,a,w)>CE(t_2,dd_2;\theta,a,w)$ if and only if $\theta<\theta(a)$.
$\Box$

\bigskip\noindent
{\bf Proof of Lemma 2:}
Fix $c=0,1,\ldots,C-1$.  From (3), the frontier $\theta_{c,c+1}(\cdot)$ between coverages $(t_c,dd_c)$ and $(t_{c+1},$ $dd_{c+1})$  is given by
\begin{eqnarray*} 
\theta_{c,c+1}(a) &=& \frac{a(t_{c+1}-t_c)}{\phi_a(dd_c)-\phi_a(dd_{c+1})} 
=    \frac{t_{c+1}-t_c}{ \int_{dd_{c+1}}^{dd_c} e^{aD} [1-H(D)] dD}.
\end{eqnarray*}
Thus, for any $c=0,1,\ldots,C-2$, $\theta_{c,c+1}(\cdot) < \theta_{c+1,c+2}(\cdot)$ on $a\in[0,\overline{a}]$ if and only if
\begin{eqnarray}
\frac{t_{c+2}-t_{c+1}}{t_{c+1}-t_c} >
\frac{ \int_{dd_{c+2}}^{dd_{c+1}} e^{aD} [1-H(D)] dD}{ \int_{dd_{c+1}}^{dd_c} e^{aD} [1-H(D)] dD}
\end{eqnarray}
for all $a\in[\underline{a},\overline{a}]$.   For any such $c$, we first show that the RHS of (A.2) decreases in $a\in(0,+\infty)$.  Adding 1 to the inverse of the RHS, it is equivalent to showing that the ratio $\int_{dd_{c+2}}^{dd_{c}} e^{aD} [1-H(D)] dD/\int_{dd_{c+2}}^{dd_{c+1}} e^{aD} [1-H(D)] dD$ is increasing in $a$, i.e., that $\int_{dd_{c+2}}^{dd} e^{aD} [1-H(D)] dD$ is log-supermodular in $(a,dd)\in(0,+\infty)\times (dd_{c+2},\overline{d})$ given $dd_{c+2}$ since $d_{c+1}<d_c$.\footnote{A positive function $f(x,y)$ is log-supermodular if $\log f(x,y)$ is supermodular.}
The latter holds by Lemma C.1 upon letting $x=a$, $y=dd$, $\overline{y}=\overline{d}$ and $y_\dagger=dd_{c+2}$.

We now show that condition (8) is necessary and sufficient for $\theta_{c,c+1}(\cdot)<\theta_{c+1,c+2}(\cdot)$ on $[\underline{a},\overline{a}]$. Because (A.2) must hold at $\underline{a}$, then (8) is necessary.  Since the RHS of (A.2) decreases in $a\in(0,+\infty)$, it is bounded above for all $a\in[\underline{a},\overline{a}]$ by the RHS of (A.2) evaluated at $\underline{a}$.  Thus (8) implies that (A.2) holds for all $a\in[\underline{a},\overline{a}]$ thereby establishing sufficiency.  $\Box$
 
\newpage
\normalsize

\begin{center}
{\bf References}
\end{center}

\smallskip\noindent
{\bf Armstrong, Mark} (1996): ``Multiproduct Nonlinear Pricing,''  {\em Econometrica}, 64, 51-75.

\smallskip\noindent
{\bf Aryal, Gaurab} and {\bf Federico Zincenko} (2023): ``Identification and Estimation of  Multidimensional Screening,''  Unpublished manuscript, University of Washington St Louis.

\smallskip\noindent
{\bf Aryal, Gaurab}, {\bf Isabelle Perrigne}, and {\bf Quang Vuong} (2016): ``Identification of Insurance Models with Multidimensional Screening,'' Unpublished manuscript, New York University.

\smallskip\noindent
{\bf Aryal, Gaurab}, {\bf Isabelle Perrigne}, and {\bf Quang Vuong} (2019):``Econometrics of Insurance Models with Multidimensional Types,'' Unpublished manuscript, New York University.

\smallskip\noindent
{\bf Aryal, Gaurab}, {\bf Isabelle Perrigne}, {\bf Quang Vuong} and {\bf Haiqing Xu}
(2024):  ``Supplement to `Econometrics of Insurance with Multidimensional Types',''
{\em Quantitative Economics Supplemental Material}.

\smallskip\noindent
{\bf Barseghyan, Levon}, {\bf Francesca Molinari}, {\bf Ted O'Donoghue} and {\bf Joshua Teitelbaum} (2013): ``The Nature of Risk Preferences: Evidence from Insurance Choices,'' {\em American Economic Review}, 103, 2499-2529.

\smallskip\noindent
{\bf Berry, Steven} and {\bf Philip Haile} (2014): ``Identification in Differentiated Product Markets,'' {\em Econometrica},  82, 1749-1797.

\smallskip\noindent
{\bf Billingsley, Patrick} (1995): {\em Probability and Measure}, 3rd Edition,  New York, USA: Academic Press.

\smallskip\noindent
{\bf Chen, Xiaohong},{\bf Matthew Gentry}, {\bf Tong Li} and {\bf Jingfeng Lu} (2023): ``Identification and Inference in First-Price Auctions with Risk-Averse Bidders and Selective Entry,''  Unpublished manuscript, Yale University.

\smallskip\noindent
{\bf Chiappori, Pierre-Andr\'e} and {\bf Bernard Salani\'e} (2000): ``Testing for Asymmetric Information in Insurance Markets,''
{\em Journal of Political Economy}, 108, 56-78.

\smallskip\noindent
{\bf Cohen, Alma} and {\bf Liran Einav} (2007): ``Estimating Risk Preferences from Deductible Choice,'' 
{\em American Economic Review}, 97, 745-788.

\smallskip\noindent
{\bf Cohen, Alma} and {\bf Peter Siegelman} (2010): ``Testing for Adverse 
Selection in Insurance Markets,'' {\em Journal of Risk and Insurance},  77, 39-84.

\smallskip\noindent
{\bf Crawford, Gregory}  and  {\bf Matthew Shum} (2007): ``Monopoly Quality Choice in Cable Television,'' {\em Journal of Law and
Economics}, 50, 181-209.

\smallskip\noindent
{\bf Cutler, David},  {\bf Amy Finkelstein}, {\bf Kathleen McGarry} (2008): ``Preference Heterogeneity and Insurance Markets: Explaining a Puzzle of Insurance,''  {\em American Economic Review, Papers and Proceedings},  98, 157-162.

\smallskip\noindent
{\bf Einav, Liran} and {\bf Amy Finkelstein} (2011): ``Selection in Insurance Markets: Theory and Empirics in Pictures,'' {\em Journal of Economic Perspectives}, 25, 115-138.

\smallskip\noindent
{\bf Fang, Hanming}, {\bf Michael Keane} and {\bf Daniel Silverman} (2008): ``Sources of Advantageous Selection: Evidence from the Medigap Insurance Market,'' {\em Journal of Political Economy}, 116, 303-350.

\smallskip\noindent
{\bf Finkelstein, Amy} and {\bf Kathleen McGarry} (2006):  ``Multiple Dimensions of Private Information: Evidence from the Long-term Care Insurance Market,'' {\em American Economic Review}, 96, 938-958.

\smallskip\noindent
{\bf Gayle, George-Levi} and {\bf Robert Miller} (2015): ``Identifying and Testing Models of Managerial Compensation,'' {\em Review of Economic Studies}, 82, 1074-1118.

\smallskip
\noindent
{\bf Gentry, Matthew} and {\bf Tong Li} (2014):  ``Identification in Auctions with Selective Entry,''
{\em Econometrica}, 82, 315-344.

\smallskip\noindent
{\bf Guerre, Emmanuel}, {\bf Isabelle Perrigne} and {\bf Quang Vuong} (2000): ``Optimal Nonparametric Estimation of First-Price Auctions,'' {\em Econometrica}, 68, 525-574.

\smallskip\noindent
{\bf Greenwood, Major} and {\bf G. Udny Yule} (1920): ``An Inquiry into the Nature of Frequency Distributions Representative of Multiple Happenings with Particular Reference to the Occurrence of Multiple Attacks of Disease or of Repeated Accidents,'' {\em Journal of the Royal Statistical Society, Series A}, 83, 255-279.

\smallskip\noindent
{\bf Hengartner, Nicolas W.} (2017): ``Adaptive Demixing in Poisson Mixture Models,''
{\em Annals of Statistics}, 25, 917-928.

\smallskip\noindent
{\bf Imbens, Guido} and {\bf Whitney Newey} (2009): ``Identification and Estimation of Triangular Simultaneous Equations Models Without Additivity,'' {\em Econometrica},  77, 1481-1512.

\smallskip\noindent
{\bf Israel, Mark} (2005): ``Services as Experience Goods: An Empirical 
Examination of Consumer Learning in Automobile Insurance,''
{\em American Economic Review}, 95, 1444-1463.

\smallskip
\noindent
{\bf Kong, Yunmi} (2018):  ``Selectice Entry in Auctions: Estimation and Evidence,''  Unpublished manuscript, Rice University.

\smallskip\noindent
{\bf Kong, Yunmi}, {\bf Isabelle Perrigne} and {\bf Quang Vuong} (2022): ``Multidimensional Auctions of Contracts: An Empirical Analysis,'' {\em American Economic Review}, 112, 1703-1736.

\smallskip\noindent
{\bf Luo, Yao}, {\bf Isabelle Perrigne} and {\bf Quang Vuong} (2017): ``Multiproduct Nonlinear Pricing: Mobile Voice Service and SMS,'' Unpublished manuscript, New York University.

\smallskip\noindent
{\bf Luo, Yao}, {\bf Isabelle Perrigne} and {\bf Quang Vuong} (2018): ``Structural Nonlinear Pricing,'' {\em Journal of Political Economy}, 126, 2523-2568.

\smallskip
\noindent
{\bf Marmer, Vadim} and  {\bf Artyom Shneyerov}  (2012):  ``Quantile Based Nonparametric Inference for First-Price Auctions,'' {\em Journal of Econometrics},  167, 345-357.

\smallskip\noindent
{\bf Matzkin, Rosa} (2003): ``Nonparametric Estimation of Nonadditive Random Functions,''
{\em Econometrica}, 71, 1339-1375.

\smallskip\noindent
{\bf Rao, B.L.S. Prakasa} (1992): {\em Identifiability in Stochastic Models: Characterization of Probability Distributions},  New York, USA: Academic Press.

\smallskip\noindent
{\bf Rochet, Jean-Charles} and {\bf Philippe Chone} (1998): ``Ironing, Sweeping and Multidimensional Screening,'' {\em Econometrica}, 66, 783-826.

\smallskip\noindent
{\bf Rochet, Jean-Charles} and {\bf Lars Stole} (2003): ``The Economics of Multidimensional Screening,'' in  {\em Advances in Eonomic Theory, 8th World Congress}, ed. by M. Dewatripont, L.P. Hansen and S. Turnovsky, Cambridge, UK: Cambridge University Press, 150-197.

\smallskip\noindent
{\bf Rothschild, Michael} and {\bf Joseph Stiglitz} (1976): ``Equilibrium in Competitive Insurance Markets: 
An Essay on the Economics of Imperfect Information,'' {\em Quarterly Journal of Economics},
90, 236-257.

\smallskip\noindent
{\bf Shohat, James} and {\bf Jacob Tamarkin} (1943): {\em The Problem of Moments}, Providence, USA: American Mathematical Society.

\smallskip\noindent
{\bf Silva, Matheus} (2024):  {\em Essays in Econometrics},  PhD dissertation, New York University. 

\smallskip\noindent
{\bf Stiglitz, Joseph} (1977): ``Monopoly, Nonlinear Pricing and Incomplete Information: The Insurance Market,'' {\em Review of Economic Studies}, 
44, 407-430.

\smallskip\noindent
{\bf Sydnor, Justin} (2010): ``(Over)Insuring Modest Risks,'' {\em American Economic Journal: Applied Economics}, 2, 177-199.

\smallskip\noindent
{\bf Talenti, Giorgio} (1987): ``Recovering a  Function from a Finite Number of Moments,'' {\em Inverse Problems}, 3.

\newpage

\begin{figure}[!h]
\centering
  \caption{$C=2$ and $C=3$ Coverages}
   \includegraphics[width=14cm, height=9cm]{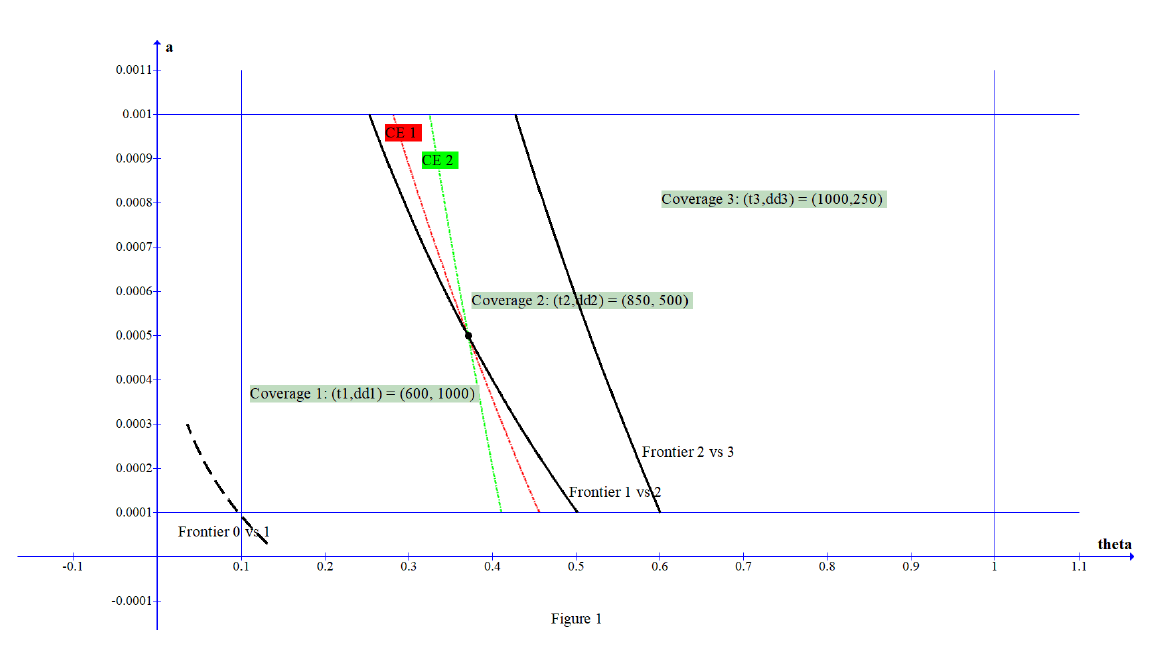}  
\label{figure1}
\end{figure}

\vspace{1pt}

\begin{figure}[!h]
\centering
  \caption{ Scatter Plot of $(a_i,\theta_i)$}
   \includegraphics[width=11cm, height=7cm]{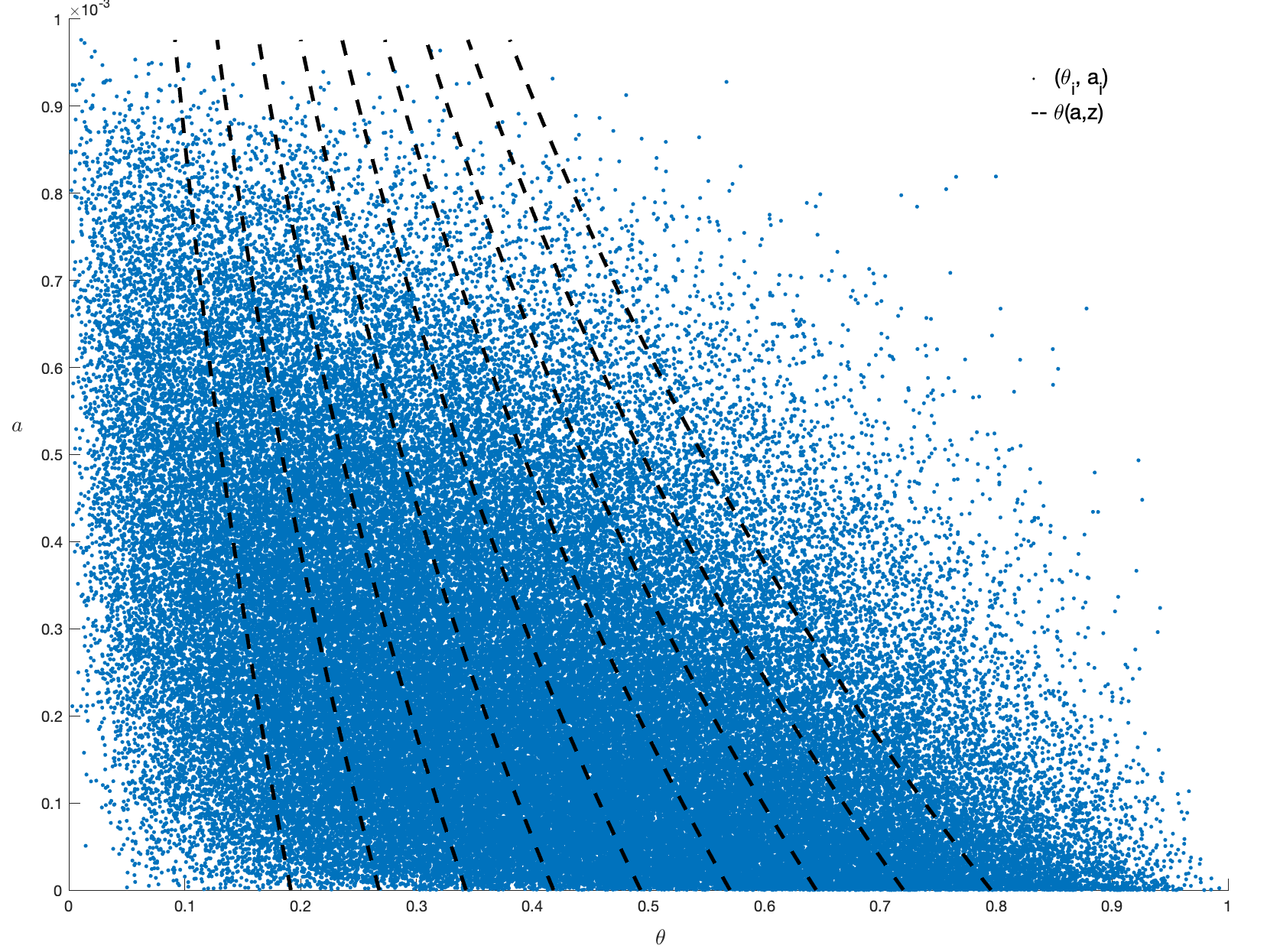}
  
\label{figure2}
\end{figure}

\newpage

\begin{figure}[!h]
\centering
  \caption{Distribution of the Number of Accidents}
   \includegraphics[width=11cm, height=7cm]{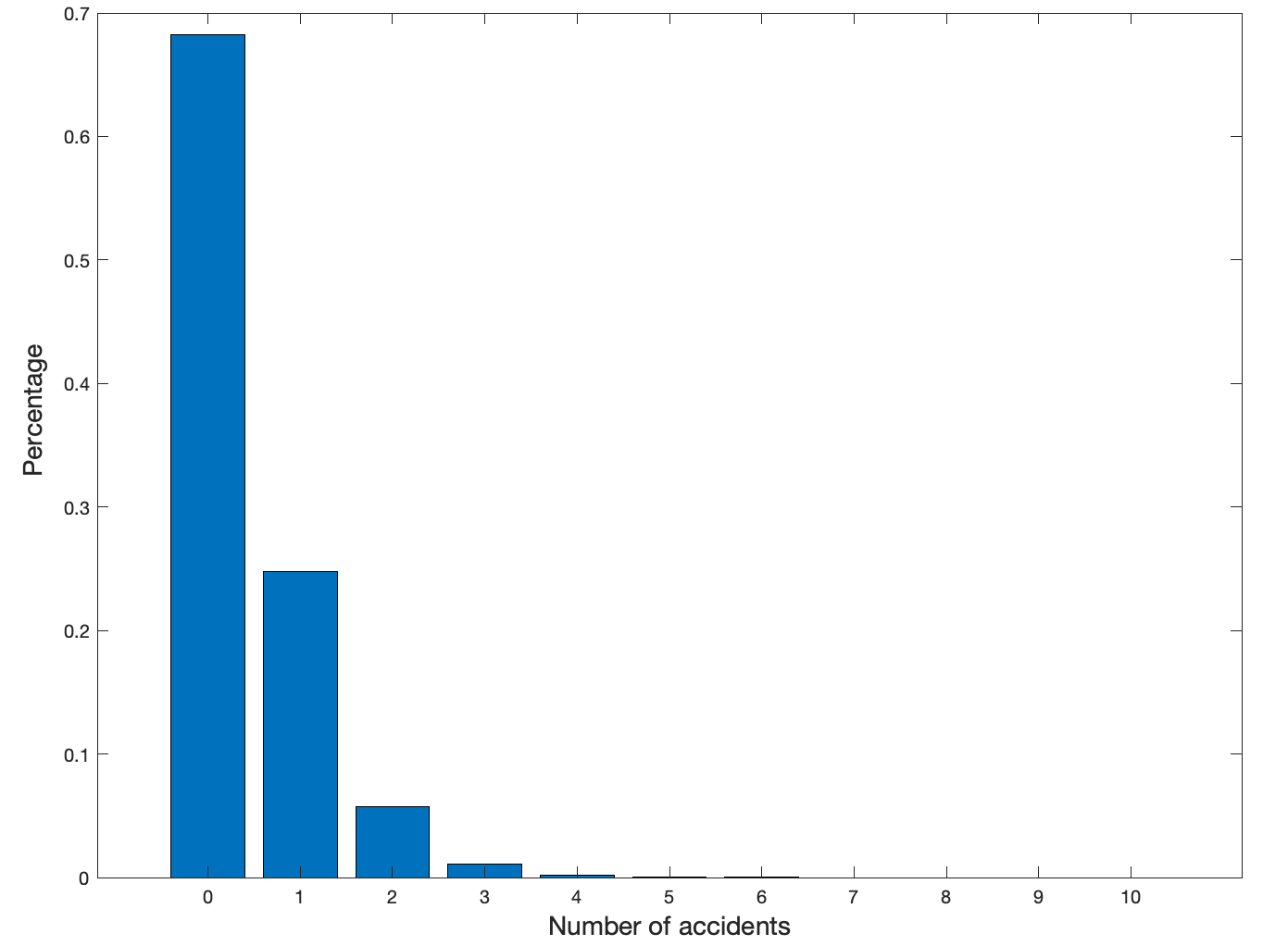}
  
\label{figure3}
\end{figure}

\vspace{1pt}

\begin{figure}[!h]
\centering
  \caption{Estimated Risk Density $\hat{f}_{\theta}(\cdot)$}
   \includegraphics[width=11cm, height=7cm]{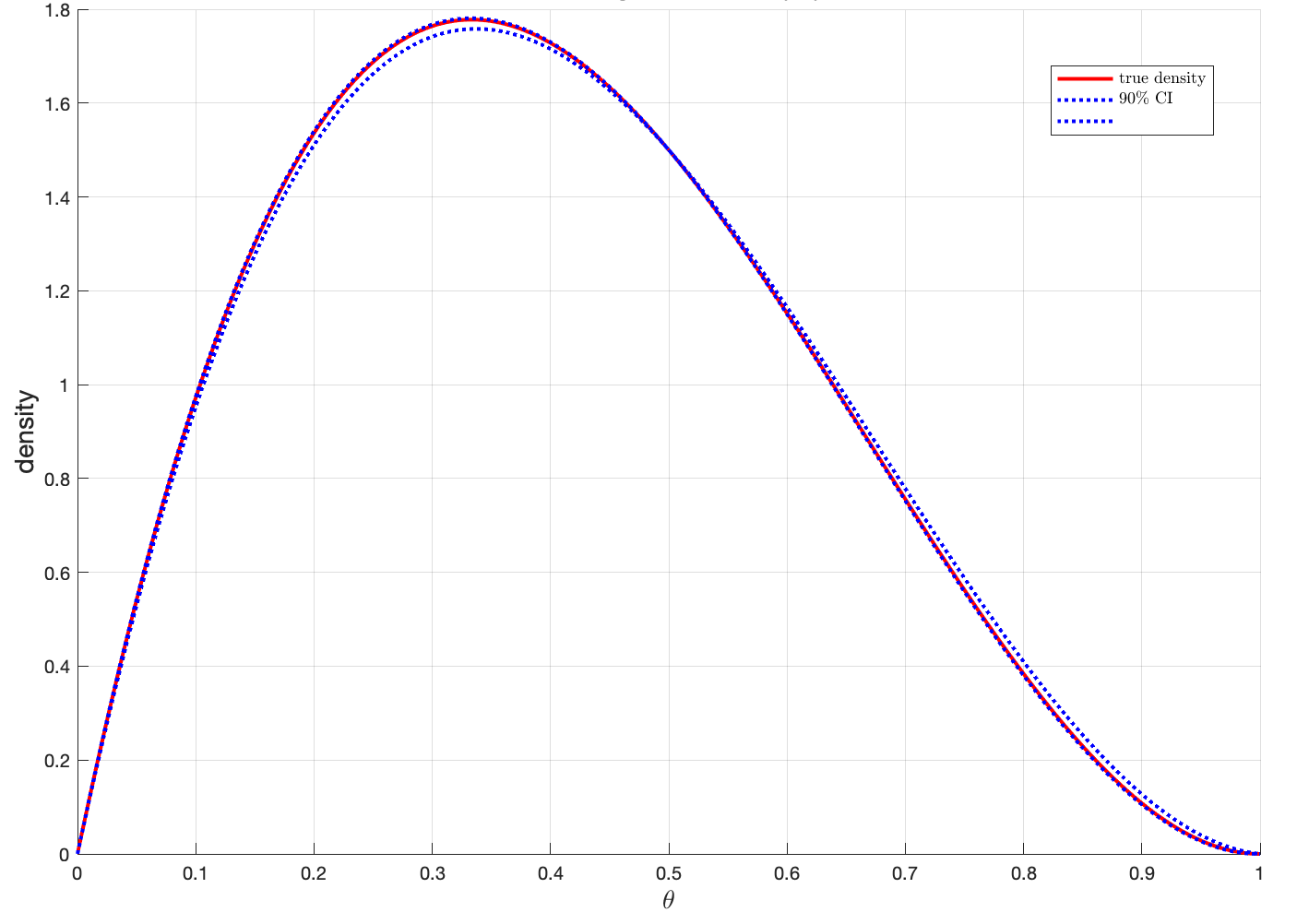}
  
\label{figure4}
\end{figure}

\newpage

\begin{figure}[!h]
\centering
\caption{ $\hat{f}_{a|\theta}(\cdot|0.4)$}
   \includegraphics[width=13cm, height=8cm]{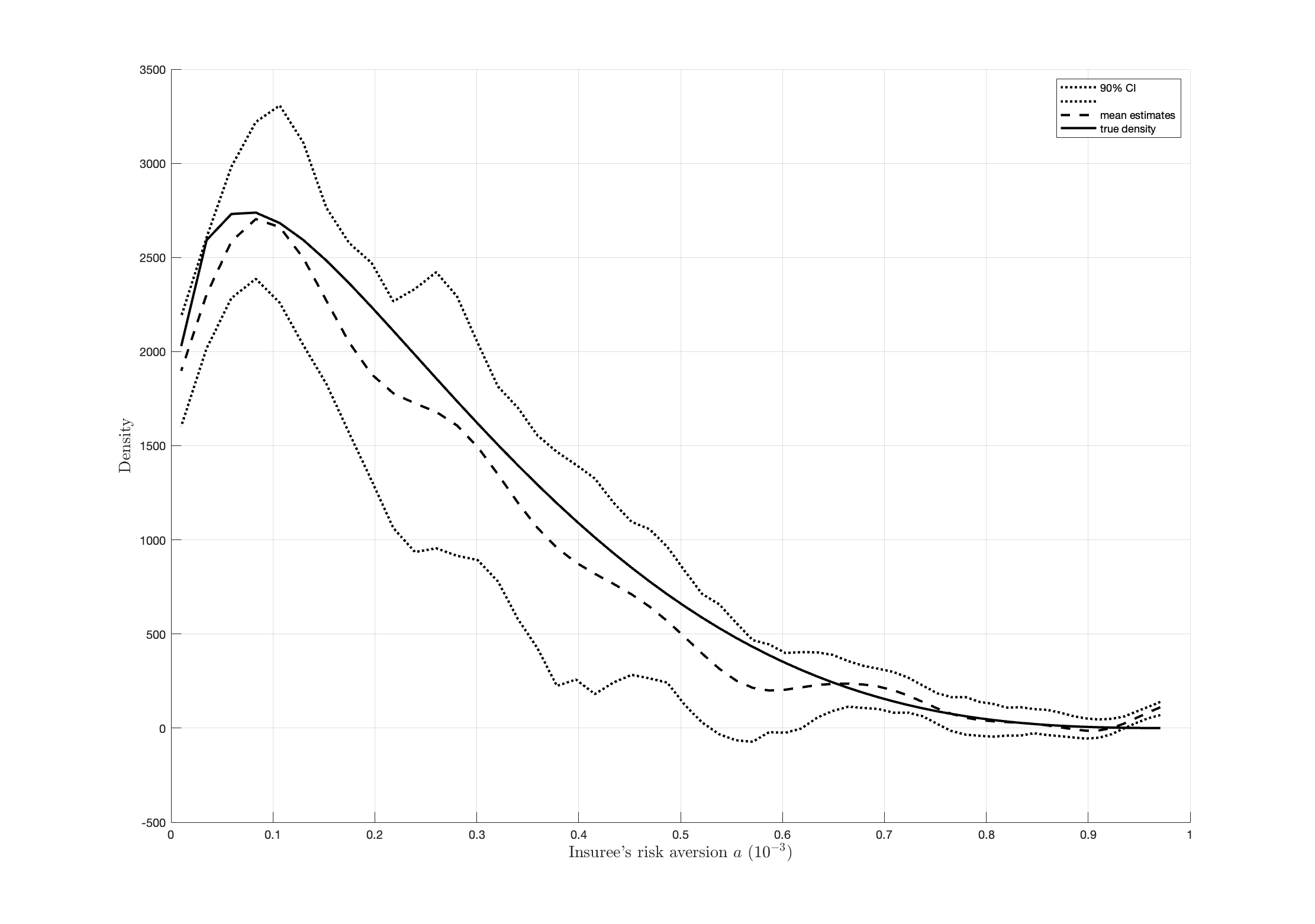}
  
\label{figure5}
\end{figure}

\vspace{1cm}

\begin{figure}[!h]
\centering
\caption{ $\hat{f}_{a|\theta}(\cdot|0.6)$}

   \includegraphics[width=13cm, height=8cm]{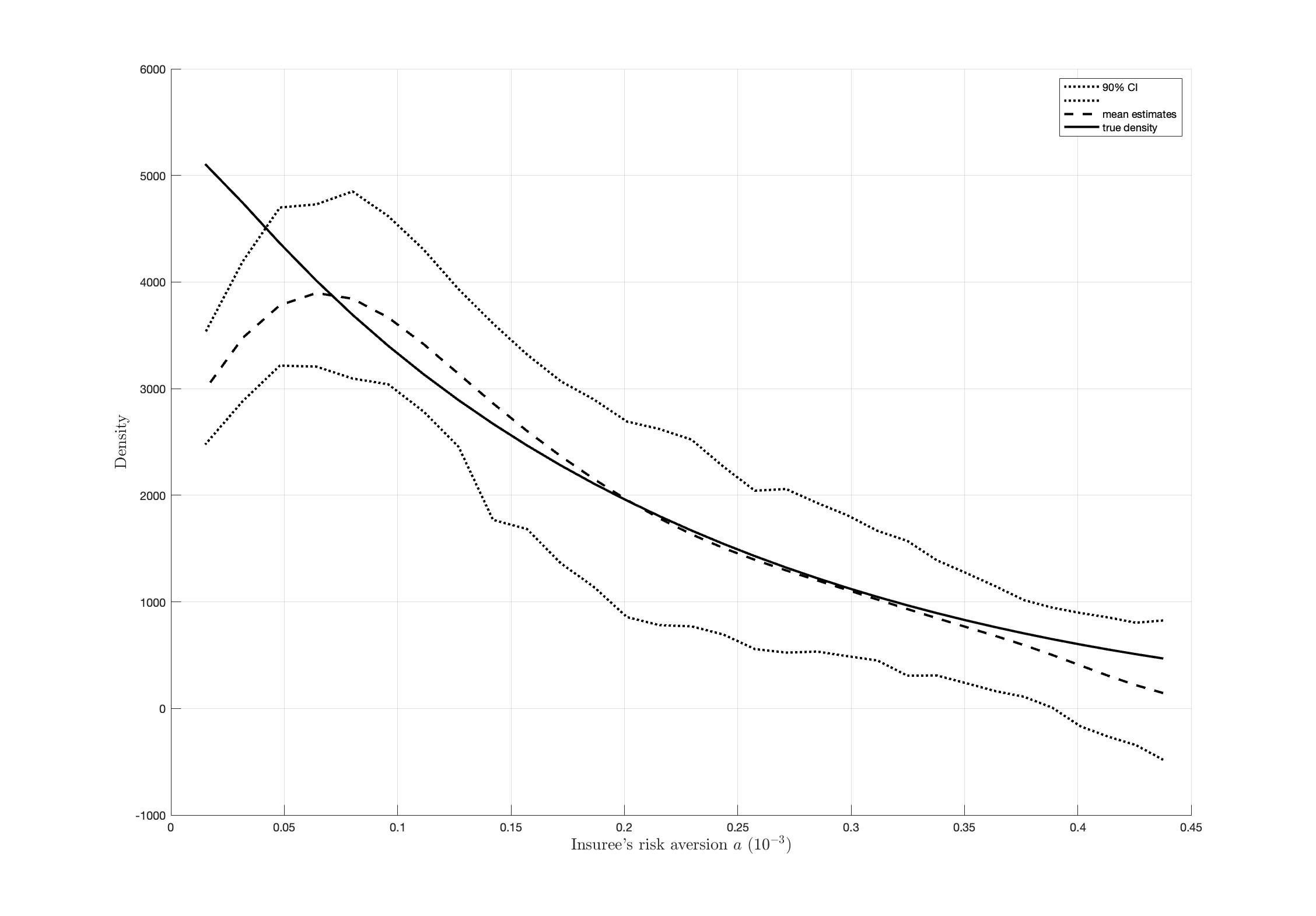}
\label{figure6}
\end{figure}
\clearpage
\newpage
\setcounter{equation}{0}
\setcounter{section}{0}

\renewcommand{\theequation}{S.\arabic{equation}}
\renewcommand{\thesection}{S.\arabic{section}}

\section{Supplmental Material}
The supplemental matertial contains three appendices.  Appendix A presents the identification results when the damage distribution is truncated at the deductible. Appendix B considers alternative specifications to the CARA function and the Poisson distribution for the insurees' utility function and the number of accidents. Appendix C establishes several lemmas mentioned in the text or used in the appendices. Some can be of independent interest whereas others are likely known though we have yet found references for them.

\bigskip
\begin{center}
{\bf Appendix A}
\end{center}

\medskip\noindent
This appendix extends the results of Section 3 when not all damages/accidents are observed because of  truncation at the deductible.
 Not observing all the accidents  limits the extent of  identification.   In particular, 
we show that $F(\cdot,\cdot|X)$ is identified up to  the knowledge  of the probability of  damage below the  lowest deductible, i.e., $H(dd_2(X)|X)$.
Thus, when one of the insurance contracts  offers  full coverage, namely $dd_2(X)=0$, this probability is known,  and the identification results of Section 3 apply despite truncation for insurees who choose $(t_1(X),dd_1(X))$.  To simplify the notation, we let $H_c(X)\equiv H(dd_c(X) |X)$, $c=1,2$. Hereafter, we assume that $0 <dd_2(X) < dd_1(X) < \overline{d}(X)$ so that $ 0< H_2(X)  < H_1(X) < 1$.  

We first derive  a relation between $1-H_1(X)$ and $1-H_2(X)$, which allows us to focus on  identification  in terms of $H_2(X)$. 
Because a claim is filed only when the damage is above the deductible,  from the observed claims, 
we identify  the damage distribution conditional on the damage being larger than the deductible, i.e., 
 the truncated damage distributions  
$H^*_c(\cdot|X)\equiv [H(\cdot|X)-H_c(X)]/[1-H_c(X))]$
on $[dd_c(X),\overline{d}(X)]$  from the insurees buying the coverage $(t_c(X),dd_c(X))$ for $c=1,2$.
Differentiating  and taking the ratio show that 
\begin{eqnarray*}
\lambda(X)\equiv \frac{h_2^*(D|X)}{h_1^*(D|X)}=\frac{1-H_1(X)}{1-H_2(X)}, 
\end{eqnarray*}  
for all  $D\geq dd_1(X)$, where $0< \lambda(X) < 1$ and $h^*_c(\cdot|X), c=1,2$ denotes the truncated damage density conditional on $X$.
In particular, the function $\lambda(\cdot)$ is identified  from the claim data, while  $H(\cdot|X)$ is identified on $[dd_2(X),\overline{d}(X)]$
up to the knowledge of $H_2(X)$.   We proceed in steps as in Section 3.

\smallskip\noindent
{\sc Identification of $F_{\theta|X}(\cdot|\cdot)$}

Let $\tilde{\theta}\equiv (1-H_2(X))\theta$ which replaces $\theta$.  We  modify our identification argument as $J$ is now unobserved. 
To identify the marginal density $f_{\tilde{\theta}|X}(\cdot|\cdot)$,  we exploit  the observed number of reported accidents $J^* \leq J$. The moment-generating function of $J^*$ given $(\chi,X)$, where $\chi\in\{1,2\}$ indicates the insuree's coverage choice, is 
\begin{eqnarray}
M_{J^*|\chi,X}(t|c,x) &=& {\rm E}[e^{J^*t}|\chi=c,X=x] \nonumber\\
&=& {\rm E}\{{\rm E}[e^{J^*t}|J,\chi,X]|\chi=c,X=x \}  \nonumber\\
&=&  {\rm E} \left\{[H_\chi(X) + (1-H_\chi(X))e^{t}]^J|\chi=c, X=x \right\} \nonumber\\
   &=& {\rm E}\left\{ {\rm E}[e^{J \log [H_\chi(X)+(1-H_\chi(X))e^t]}|\theta,a,\chi,X] |\chi=c,X=x   \right\} \nonumber\\
  &=&  {\rm E} \left[e^{\theta[H_\chi(X) + (1-H_\chi(X))e^t-1]}|\chi=c,X=x  \right] \nonumber\\
  &=& M_{\theta|\chi,X} [(1-H_\chi(X))(e^t-1)|c,x],
\end{eqnarray}
where the third equality uses the moment-generating function of $J^*$ given $(J,\chi,X)$, which is a Binomial  ${\cal B}(J,1-H_\chi(X))$ 
from A2-(iv),  and the fifth equality 
follows from A2-(iii) and  the moment-generating function of the Poisson distribution. 
Thus, 
\begin{eqnarray*}
M_{\theta|\chi,X}[u|c,x] = M_{J^*|\chi,X} \left[\log \left(1 + \frac{u}{1-H_\chi(X)}   \right)\Big{|}c,x   \right],
\end{eqnarray*}
for $u\in (-1+H_\chi(X),+\infty)$, where $H_{\chi}(X) <1$.
Hence the distribution of risk $\theta$ given $(\chi,X)$ is identified up to the knowledge of $H_\chi(X)$. 

Since $\tilde{\theta}=(1-H_2(X)) \theta$, its moment-generating function given $(\chi,X)$ is 
\begin{eqnarray}
M_{\tilde{\theta}|\chi,X}(u|c,x) &=& M_{\theta|\chi,X}(u(1-H_2(x))|c,x) \nonumber \\
&=& \left\{ \begin{array}{ll} M_{J^*|\chi,X}\left[\log \left(1 + \frac{u}{\lambda(x)} \right)|1,x \right]  & \ {\rm if} \ c=1, \\
                              M_{J^*|\chi,X}\left[\log \left(1 + u \right)|2,x \right]                   &\ {\rm if} \ c=2,
\end{array} \right. 
\end{eqnarray}
for all $u \in (-\lambda(x),+\infty)$ and $u\in (-1,+\infty)$, respectively, where $\lambda(x) >0$.
Thus, the moment-generating function of $\tilde{\theta}$ given $X$ is the weighted average
\begin{eqnarray*}
M_{\tilde{\theta}|X}(u|x)\! &=& \!{\rm E} \{{\rm E}[e^{u \tilde{\theta}}|\chi,X]|X=x \}\nonumber \\
  \!&=&\! M_{J^*|\chi,X}\! \left[\log \!\left(1\!+\! \frac{u}{\lambda(x)}\right) |1,x  \right]\! \nu_1(x)    
        + M_{J^*|\chi,X} \!\left[\log \left(1\!+\! u \right) |2,x  \right]\! \nu_2(x),
\end{eqnarray*}
for $u \in (-\lambda(x),+\infty)$. Thus $f_{\tilde{\theta}|X}(\cdot|\cdot)$ is identified as $\lambda(X)$ is identified while  $\nu_1(X)$ and $\nu_2(X)$ are the proportions of insurees choosing coverages 1 and 2 that are observed in the data. Since $f_{\theta|X}(\theta|x)= (1-H_2(x)) f_{\tilde{\theta}|X}((1-H_2(x))\theta|X)$,
the conditional distribution $F_{\theta|X}(\cdot|x)$ is identified up to $H_2(x)$.

\smallskip\noindent
{\sc Identification of $F_{\theta|X}(\cdot|\cdot)$}

Let $\tilde{\theta}\equiv (1-H_2(X))\theta$ which replaces $\theta$.  We  modify our identification argument as $J$ is now unobserved. 
To identify the marginal density $f_{\tilde{\theta}|X}(\cdot|\cdot)$,  we exploit  the observed number of reported accidents $J^*$. The 
moment-generating function of $J^*$ given $(\chi,X)$, where $\chi\in\{1,2\}$ indicates the insuree's coverage choice, is 
\begin{eqnarray}
M_{J^*|\chi,X}(t|c,x) &=& {\rm E}[e^{J^*t}|\chi=c,X=x] \nonumber\\
&=& {\rm E}\{{\rm E}[e^{J^*t}|J,\chi,X]|\chi=c,X=x \}  \nonumber\\
&=&  {\rm E} \left\{[H_\chi(X) + (1-H_\chi(X))e^{t}]^J|\chi=c, X=x \right\} \nonumber\\
   &=& {\rm E}\left\{ {\rm E}[e^{J \log [H_\chi(X)+(1-H_\chi(X))e^t]}|\theta,a,\chi,X] |\chi=c,X=x   \right\} \nonumber\\
  &=&  {\rm E} \left[e^{\theta[H_\chi(X) + (1-H_\chi(X))e^t-1]}|\chi=c,X=x  \right] \nonumber\\
  &=& M_{\theta|\chi,X} [(1-H_\chi(X))(e^t-1)|c,x],
\end{eqnarray}
where the third equality uses the moment-generating function of $J^*$ given $(J,\chi,X)$, which is a Binomial  ${\cal B}(J,1-H_\chi(X))$ 
from A2-(iv),  and the fifth equality 
follows from A2-(iii) and  the moment-generating function of the Poisson distribution. 
Thus, 
\begin{eqnarray*}
M_{\theta|\chi,X}[u|c,x] = M_{J^*|\chi,X} \left[\log \left(1 + \frac{u}{1-H_\chi(X)}   \right)\Big{|}c,x   \right],
\end{eqnarray*}
for $u\in (-1+H_\chi(X),+\infty)$, where $H_{\chi}(X) <1$.
Hence the distribution of risk $\theta$ given $(\chi,X)$ is identified up to the knowledge of $H_\chi(X)$. 

Since $\tilde{\theta}=(1-H_2(X)) \theta$, its moment-generating function given $(\chi,X)$ is 
\begin{eqnarray}
M_{\tilde{\theta}|\chi,X}(u|c,x) &=& M_{\theta|\chi,X}(u(1-H_2(x))|c,x) \nonumber \\
&=& \left\{ \begin{array}{ll} M_{J^*|\chi,X}\left[\log \left(1 + \frac{u}{\lambda(x)} \right)|1,x \right]  & \ {\rm if} \ c=1, \\
                              M_{J^*|\chi,X}\left[\log \left(1 + u \right)|2,x \right]                   &\ {\rm if} \ c=2,
\end{array} \right. 
\end{eqnarray}
for all $u \in (-\lambda(x),+\infty)$ and $u\in (-1,+\infty)$, respectively, where $\lambda(x) >0$.
Thus, the moment generating function of $\tilde{\theta}$ given $X$ is the weighted average
\begin{eqnarray*}
M_{\tilde{\theta}|X}(u|x)\! &=& \!{\rm E} \{{\rm E}[e^{u \tilde{\theta}}|\chi,X]|X=x \}\nonumber \\
  \!&=&\! M_{J^*|\chi,X}\! \left[\log \!\left(1\!+\! \frac{u}{\lambda(x)}\right) |1,x  \right]\! \nu_1(x)    
        + M_{J^*|\chi,X} \!\left[\log \left(1\!+\! u \right) |2,x  \right]\! \nu_2(x),
\end{eqnarray*}
for $u \in (-\lambda(x),+\infty)$. Thus $f_{\tilde{\theta}|X}(\cdot|\cdot)$ is identified as $\lambda(X)$ is identified while  $\nu_1(X)$ and $\nu_2(X)$ are the proportions of insurees choosing coverages 1 and 2 that are observed in the data. Since $f_{\theta|X}(\theta|x)= (1-H_2(x)) f_{\tilde{\theta}|X}((1-H_2(x))\theta|X)$,
the conditional distribution $F_{\theta|X}(\cdot|x)$ is identified up to $H_2(x)$.
 
\smallskip\noindent
{\sc Identification of $F(\theta,a|X)$}

Following Section 3, we consider  the probability that a $(\theta,a)$-insuree with characteristics $X$ chooses the coverage $(t_1(X),dd_1(X))$.
Using (5) and $1-H(D|X)=(1-H_2(X))(1-H^*_2(D|X))$, we remark that the indifference frontier between the two coverages in the space $(\tilde{\theta},a)$ is given by
\begin{eqnarray*}
\tilde{\theta}(a,X) = \frac{t_2(X)-t_1(X)}{\int_{dd_2(X)}^{dd_1(X)} e^{aD} [1-H^*_2(D|X)] dD},
\end{eqnarray*} 
leading to the inverse $a(\tilde{\theta},X)$, which is identified. From Bayes' rule, as in (6), we have 
\begin{eqnarray*}
F_{a|\tilde{\theta},X}(a(\tilde{\theta},x)|\tilde{\theta},x) = \frac{f_{\tilde{\theta}|\chi,X}(\tilde{\theta}|1,x) \nu_1(x) }
         {f_{\tilde{\theta}|X}(\tilde{\theta}|x)},
\end{eqnarray*}
where $\nu_1(x)$  is observed and  $f_{\tilde{\theta}|X}(\tilde{\theta}|x)$ is identified from the first step. 
Moreover, $f_{\tilde{\theta}|\chi,X} (\cdot|1,x)$  is identified because its moment-generating function
$M_{\tilde{\theta}|\chi,X}(\cdot|1,x)$ is identified on $(-\lambda(x,),$ $+\infty)$ as shown in (S.2). Thus  $F_{a|\tilde{\theta},X}(\cdot|\tilde{\theta},x)$
is identified on the frontier $a(\tilde{\theta},x)$.

Lastly, we note that $F_{a|\tilde{\theta},X}(a(\tilde{\theta},x)|\tilde{\theta},x)= F_{a|\theta,X}(a(\theta,x)|\theta,x)$
thereby identifying the latter up to $H_2(x)$ since $\tilde{\theta}= (1-H_2(x))\theta$.
Hence under the exclusion restriction and full support condition in  A3, $F_{a|\theta,X}(\cdot|\cdot,\cdot)$ and hence $F(\theta,a|X,Z)$   are identified up to the knowledge of $H_2(X)$.  This result is formally stated next.

\medskip\noindent
{\bf Proposition 2:} {\em   Suppose that there are two offered  coverages with $0 < dd_2(X) < dd_1(X) < \overline{d}(X)$, and  damages are observed only when they are  above the deductible for each insuree. Under A2--A3, the structure $[F(\cdot,\cdot|X),H(\cdot|X)]$ is identified up to $H_2(X)$.}

\medskip\noindent
In the absence of the full support condition A3-(ii), the comments after Proposition 1 still apply, up to the knowledge of $H_2(X)$.
We note that having a larger number of coverages $C>2$ can only improve the identification results for two reasons. First, as in Section 3,  more frontiers
of the form (5) are more likely to cover the whole support $\Theta(x) \times {\cal A}(x_0)$ when $Z$ varies. Second, the lowest deductible $dd_C(x)$ among the  $C$ coverages might not be binding  as the probability of  damage below $dd_C(x)$ is small.  In practice, this can be assessed by estimating the conditional damage distribution  in the neighborhood $[dd_C(X),(1+\delta) dd_C(X)]$ with $\delta$ small.

A consequence of Proposition 2 is that the structure $[F(\cdot,\cdot|X), H(\cdot|X)]$ is identified if and only if $H_2(X)$ is identified.
The next result shows that $H_2(X)$ is not identified.

\medskip\noindent
{\bf Proposition 3:} {\em Under  the conditions of Proposition 2,  $H_2(X)$ is not identified.}

\medskip\noindent
The proof relies on exhibiting an observationally equivalent structure, as shown below.

\noindent
{\bf Proof of Proposition 3:} Given Proposition 2, $H_2(X)$ is identified if and only if the structure $[F(\cdot,\cdot|X),H(\cdot|X)]$ is. Thus, it suffices to show that the latter is not identified.  Let  $[F(\cdot,\cdot|X),H(\cdot|X)]$ be a structure  satisfying A2--A3.
We construct  a second structure $[\tilde{F}(\cdot,\cdot|X),$  $\tilde{H}(\cdot|X)]$ as follows. 
Let $\kappa(\cdot)$ be a (measurable) function of $X$ that can be arbitrarily large as long as $\kappa(x) \geq 1-H_2(x)$ for all $x\in {\cal S}_X$.  Let $\tilde{\theta}=\kappa(X) \theta$ and $\tilde{a}=a$. Since $\kappa(X)>0$ from $1-H_2(X)>0$, we have 
\begin{eqnarray*}
\tilde{f}(\cdot,\cdot|X)= \frac{1}{\kappa(X)} f\left(\frac{\cdot}{\kappa(X)},\cdot\mid X\right).
\end{eqnarray*} 
Let $\tilde{h}(D|X)= [1/\kappa(X)] h(D|X)$ for $D\geq dd_2(X)$. 
Thus, $\int_{dd_2(X)}^{\overline{d}(X)} \tilde{h}(D|X) dD = [1-H_2(X)]/\kappa(X)$  $\in (0,1]$.  For $D \in [0, dd_2(X))$, define $\tilde{h}(D|X)$ to be nonnegative as long as
$\int_0^{dd_2(X)} \tilde{h}(D|X) dD = 1- [1-H_2(X)]/\kappa(X)$.
In particular,  upon evaluating $\tilde{H}(D|X)$,  it is easy to verify that 
\begin{eqnarray}
1- \tilde{H}(D|X) = \frac{1}{\kappa(X)}[1-H(D|X)]
\end{eqnarray}
for $D \geq dd_2(X)$.  Moreover, the second structure $[\tilde{F}(\cdot,\cdot|X), \tilde{H}(\cdot|X)]$ satisfies  A2--A3.

We now show that these two structures are observationally equivalent, i.e., they lead to the same distribution for the observables 
$( J^*, D^*_1,\ldots,D^*_{J^*}, \chi)$ given $X$ and $(t_1, dd_1, t_2, dd_2)$, where $J^*$ and $D^*$ refer to the number of reported accidents
and their corresponding damages, respectively, while $\chi$ indicates the coverage  chosen by the insuree.
Regarding the distribution of $\tilde{\chi}$ given $X$, we note that $\tilde{\chi}=\chi$. The latter follows from  $\tilde{\chi}=1$ if and only if 
$(\tilde{\theta},a) \in \tilde{{\cal C}}_1(X)$, i.e., $\tilde{\theta} \leq \tilde{\theta}(a,X)$.
But $\tilde{\theta}=\kappa \theta$ while the  frontier (5)  defining the insurees' coverage choice for the second structure satisfies
\begin{eqnarray*} 
\tilde{\theta}(a,X)&=&  \frac{t_2(X)-t_1(X)}{ \int_{dd_2(X)}^{dd_1(X)} e^{aD} (1-\tilde{H}(D|X)) dD} 
                  = \frac{t_2(X)-t_1(X)}{ \int_{dd_2(X)}^{dd_1(X)} e^{aD} \frac{1}{\kappa(X)}(1- H(D|X)) dD}\\
                  &=& \kappa(X) \theta(a,X)
\end{eqnarray*}
using (S.3). Hence  $\tilde{\theta} \leq \tilde{\theta}(a,X)$ is equivalent to  $\theta \leq \theta(a,X)$.  That is, $\tilde{\chi}=1$ if and only if $\chi=1$. Thus, the distributions of $\tilde{\chi}$ and $\chi$ given $X$ are the same, i.e., 
$\tilde{\nu}_c(X)= \nu_c(X)$ for $c=1,2$.

Regarding the distribution of $\tilde{J}^*$ given $(\tilde{\chi},\!X)\!=\!(\chi,\!X)$, from (S.1), its moment-generating function is
\begin{eqnarray*}
M_{\tilde{\theta}|\chi,X} [(1-\tilde{H}_\chi(X))(e^t-1)|c,x] &=& M_{\theta|\chi,X}[(1-H_\chi(X))(e^t-1)|c,x] \\
                                                                 &=& M_{J^*|\chi,X}[t|c,x],
\end{eqnarray*}   
where the first equality uses $M_{\tilde{\theta}|\chi,X}(u|c,x)= M_{\theta|\chi,X}[\kappa(x) u|c,x]$, which follows from $\tilde{\theta}=\kappa(X)\theta$, and $1-\tilde{H}_c(X)= (1-H_c(X))/\kappa(X)$, which follows from (S.3).
Hence, the distribution of $\tilde{J}^*$ given $(\chi,X)$ is the same as that of $J^*$ given $(\chi,X)$.

Lastly, regarding the distribution of reported damage $\tilde{D}^*$ given $(\tilde{J}^*,\chi,X)$ we have
\begin{eqnarray*}
\tilde{H}^*_\chi(\cdot|X) = \frac{\tilde{H}(\cdot|X) - \tilde{H}_\chi(X)}{1-\tilde{H}_\chi(X)} = 
\frac{H(\cdot|X) - H_\chi(X)}{1-  H_\chi(X)} = H^*_\chi(\cdot|X)
\end{eqnarray*} 
on $[dd_2(X),\overline{d}(X)]$ where we used $1-\tilde{H}_\chi(\cdot|X)= (1-H_\chi(\cdot|X))/\kappa(X)$ and (S.3). Hence, the two structures lead to the same distribution for the observables as desired. $\Box$

\medskip\noindent
Proposition 3 shows that all the information provided by the model and the data have been exhausted.
The nonidentification  can be explained  as follows. It arises from a compensation between the increase (decrease) in the number of accidents and an appropriate decrease (increase) in the probability of damages  greater than the deductible.
From the insuree's perspective, such  compensation maintains the relative ranking between the two contracts. Thus, if a  $(\theta,a)$-insuree 
buys $(t_1(X),dd_1(X))$, then  the $((1-H_2(X))\theta,a)$-insuree  also buys the same coverage if there is an appropriate increase  in the probability of damages being greater than $dd_1(X)$. From the insurer's perspective,  the decrease  in the average number of accidents is compensated by an appropriate increase in the probability that the damage is above the deductible so that the expected payment to the insuree  remains the same under either coverage.

\smallskip\noindent
{\sc Identification Strategies for $H_2(X)$} 

We discuss identification strategies for the probability $H_2(X)$.  
Any  strategy that identifies $H_2(X)$  identifies the structure $[F(\cdot,\cdot|X),$ $H(\cdot|X)]$. 
A first  strategy is to parameterize the damage distribution $H(\cdot|X)$ as $H(\cdot|X;\beta)$ on $[0,\overline{d}(X)]$ with $\beta \in {\cal B} \subset \Real^q$.  Observations on reported damages $D^*$ identify $\beta$ and hence $H(\cdot|X)$ on $[0, \overline{d}(X)]$.
Thus $H_2(X)\equiv H(dd_2(X)|X;\beta)$ is identified.
In particular, we can choose a parametrization to fit the estimated truncated damage distribution $H^*(\cdot|X)$ on $[dd_2(X),\overline{d}(X)]$.

A second strategy is to consider additional data  on the average  number of accidents.
For instance,   suppose that for every $x \in {\cal S}_X$, we know the average number of  accidents   $\mu(x)\equiv {\rm E}[J|X=x]={\rm E}\{{\rm E}[J|\theta,X=x]|X=x\}={\rm E}[\theta|X=x]$ by A2-(iii). For 
 the average number of reported accidents, we have  $\mu^*_c(x)\equiv {\rm E}[J^*|\chi=c, X=x]={\rm E}\{ {\rm E}[J^*|J,\chi=c,X=x]|\chi=c,X=x \}=
{\rm E}[J (1-H_c(X))|\chi=c,X=x]=[1-H_c(x)]
{\rm E}[\theta|\chi=c,X=x]$ for $c=1,2$ since $J^*$ given $(J,\chi,X)$ is distributed as a Binomial with parameters $(J,1-H_{\chi}(X))$.
Thus
\begin{eqnarray*}
\mu(x)&=&  \nu_1(x) {\rm E}[\theta|\chi=1,X=x] + \nu_2(x) {\rm E}[\theta|\chi=2,X=x]\\ 
        &=&\frac{1}{1-H_2(x)} \left( \nu_1(x) \frac{\mu^*_1(x)}{\lambda(x)} + \nu_2(x) \mu^*_2(x) \right),
\end{eqnarray*}
identifying $H_2(x)$, given that $\nu_c(x)$, $\mu^*_c(x), c=1,2$, and $\lambda(x)$ are identified from the data. 
Alternatively, suppose  we know only ${\rm E}[J|X=x_*]$ for some $x_*$.
Using the same argument establishes the identification of $H_2(x_*)$. This  result combined with a support assumption such as
$\overline{\theta}(x)=\overline{\theta}$ for every $x$ identifies $H_2(x)$. To see this, note that 
$\overline{\tilde{\theta}}(x) = (1-H_2(x)) \overline{\theta}(x)$, where $\overline{\tilde{\theta}}(x)$ is the upper boundary 
of the support of $f_{\tilde{\theta}|X}(\cdot|X=x)$, which is identified. Applying this equation at $x_*$ identifies 
$\overline{\theta}=\overline{\theta}(x_*)=\overline{\tilde{\theta}}(x_*)/(1-H_2(x_*))$. Applying  this equation again at different values $x$ identifies $H_2(x)$.
A similar argument applies at the lower bound $\underline{\theta}(x)=\underline{\theta}$.\footnote{In contrast, information on the average damage is not sufficient. We note that
$
{\rm E}(D|X=x) =  H_2(x){\rm E}[D|D \leq dd_2(x),X=x]  
    + (1-H_2(x)) {\rm E}[D|D \geq dd_2(x),X=x]$, 
where ${\rm E}[D|D \geq dd_2(x),X=x]$ is identified from the data.
Thus,  identification of $H_2(x)$ requires  to know both ${\rm E}[D|D \leq dd_2(x),X=x]$ and
${\rm E}(D|X=x)$.}

A third strategy is to derive sharp bounds on the probability $H_2(X)$.
This approach, also known as partial identification, was developed by  e.g. Manski and Tamer (2002), Haile and Tamer (2003) and Chernozhukov, Hong and Tamer (2007).  Our bounds are nonparametric.
Let  $[F(\cdot,\cdot|X), H(\cdot|X)]$ be the structure generating the observables. 
Fix a value $x \in {\cal S}_X$. Proposition 2 implies that it is sufficient to determine the identified set for $H_2(x)$, i.e., the set of values $\tilde{H}_2(x)$ that are observationally equivalent to $H_2(x)$. To be precise, this is the set of values $\tilde{H}_2(x)$ corresponding to  structures 
$[\tilde{F}(\cdot,\cdot|X),\tilde{H}(\cdot|X)]$ that are observationally equivalent to $[F(\cdot,\cdot|X), H(\cdot|X)]$. Indeed, Proposition 2 shows that $[F(\cdot,\cdot|X),H(\cdot|X)]$ is identified up to $H_2(\cdot)$.
The proof of Proposition 3 above  shows that any value $\tilde{H}_2(X) = 1 - (1/\kappa)[1-H_2(x)]$ for $\kappa \in  (1-H_2(X),\infty)$ is observationally 
equivalent to $H_2(x)$. Thus, the   identified set for $H_2(x)$ is $(0,1)$.\footnote{We can then obtain sharp bounds for the structure $[F(\cdot,\cdot|\cdot),H(\cdot|\cdot)]$. Fixing $X=x$, let $H_2(x)=h \in (0,1)$ leading to a unique structure $[F_h(\cdot,\cdot|x),H_h(\cdot|x)]$ by Proposition 2.  The identified set for $[F(\cdot,\cdot|x),H(\cdot|x)]$ is the collection of such $[F_h(\cdot,\cdot|x),H_h(\cdot|x)]$ when $h$ runs over $(0,1)$. To derive bounds, we can follow Haile and Tamer (2003) by taking lower and upper envelopes for each structural distribution $F(\cdot,\cdot|x)$ or $H(\cdot|x)$.}

To tighten the  upper bound, we can use 
some empirical evidence.  From Cohen and Einav (2007),
the estimated damage density  decreases when the damage approaches the deductible from above,  suggesting that the density below the deductible is not greater than its value at the deductible. Thus we can assume that  the damage density satisfies
$h(D|x)\leq h[dd_2(x)|x]$ 
for every $D\leq dd_2(x)$ and $x$.
Integrating both sides from $0$ to $dd_2(x)$ we obtain $0\leq H_2(x) \leq dd_2(x) h(dd_2(x)|x)$. 
Dividing both sides  by $1-H_2(x)$, and using the definition of the truncated density $h_2^*(\cdot|x)$, we obtain
\begin{eqnarray*} 
0\leq \frac{H_2(x)}{1-H_2(x)}\leq dd_2(x) h_2^*(dd_2(x)|x).
\end{eqnarray*}
Solving for $H_2(x)$ gives the upper bound
\begin{eqnarray*} 
 H_2(x)\leq \frac{dd_2(x)h_2^*(dd_2(x)|x)}{1+dd_2(x)h_2^*(dd_2(x)|x)}\equiv \overline{H}_2(x),
\end{eqnarray*}
which is strictly less than 1. Thus the identified set for $H_2(x)$ reduces to 
$(0,\overline{H}_2(x)]$.   The upper bound  can be estimated from 
observables.\footnote{The comment in the previous footnote applies with $h$ running over
$(0,\overline{H}_2(x)]$.}

The extension of Section 3 applies  up to the identification of $H_2(\dot)$. It is worth noting that even if damages were not observed below the deductibles, condition (8) would still be implementable and verifiable.  Indeed, upon dividing by $[1-H(dd_C)]$ the ratio in the RHS is equal to $\int_{dd_{c+2}}^{dd_{c+1}} [1-H^*_C(D)] dD / 
\int_{dd_{c+1}}^{dd_{c}} [1-H^*_C(D)] dD$ where $H^*_C(\cdot)=[H(\cdot)-H(dd_C)]/[1-H(dd_C)]$ is the distribution of $D$ conditional on $D>dd_C$.  The latter distribution is identified from the claims of individuals buying the highest coverage $(t_C,dd_C)$, i.e., the lowest deductible $dd_C$. Thus  the  Corollary still applies, and one should observe that the observed coverages should lie on a convex curve in the 
$(t,dd)$-space.
The estimation procedure of Section 4 extends to this case with some appropriate adjustments, and the analyst can choose an estimator for $H_2(\cdot)$ in line with the identification strategies discussed above.

\vspace{0.8cm}

\begin{center}
{\bf Appendix B}
\end{center}

\noindent
This appendix extends the results of Section 3 under alternative specifications to the CARA function and the Poisson distribution for the insurees' utility function and  the number of accidents. 
As before, we assume   that heterogeneity across insurees is aracterized by a bidimensional vector $(\theta,a)$ of private information. Thus, by necessity parametric assumptions on the utility function and the distribution of the number of accidents follow.  Specifically, we let the increasing and concave  utility function $U(\cdot;a)$ be parameterized by $a \in \Real_+$ capturing the  insuree's risk aversion, while the distribution of the number of accidents $P(\cdot;\theta)$ is parameterized by  $\theta$ capturing  the insuree's risk. These need no longer be the CARA utility function nor the Poisson distribution. We stress that the CARA-Poisson specification is widely used for its mathematical tractability that we  loose under alternative specifications.  We still leave the joint distribution of $(\theta,a)$ unspecified.  Thus  our approach remains semiparametric as we consider nonparametric mixtures  for the utility function and the distribution of the number of accidents in the population.

We extend A1 by replacing A1-(i,iii) with more general assumptions on the utility function and the distribution of the number $J$ of accidents.

\medskip\noindent
{\bf Assumption A1':} {\em  While A1-(ii,iv) remain, A1-(i,iii) are replaced by \\
(i)  The insuree's utility function $U(x;a)$ is increasing and concave in $x$, where   $U(\cdot;a')$ is more risk averse than $U(\cdot;a)$ whenever $a'>a$,\footnote{ By definition, this means that there exists an  increasing and  concave function $q(\cdot)$ possibly depending on $(a',a)$ such that $U(\cdot;a')= q[U(\cdot;a)]$.} \\
(iii) The number of accidents $J$ given $\theta$ is distributed as $P(\cdot;\theta)$, where $P(\cdot;\theta')$ First-Order Stochastically Dominates  (FOSD) $P(\cdot;\theta)$, i.e., $P(\cdot;\theta')\stackrel{FOSD}{\succ} P(\cdot;\theta)$, whenever $\theta'>\theta$.
}

\medskip\noindent
The main role of A1'-(i,iii) is to rank insurees in terms of their risk aversion and risk parameters $a$ and $\theta$, respectively.  Beside the CARA utility function, a well-known utility function satisfying A1'-(i) is the Constant Relative Risk Aversion (CRRA) utility function $x^{1-a}/[1-a]$ for $a\in(0,1)$. More generally, we can consider  the Hyperbolic Absolute  Risk Aversion (HARA)  utility function which  nests  ARA and RRA as special cases.  It is defined as
\begin{eqnarray*}
U(x;\alpha,\beta,\gamma) = \frac{1-\alpha}{\alpha} \left(  \frac{\gamma x}{1-\alpha} +\beta \right)^\alpha,
\end{eqnarray*}
where $\alpha\neq 0$, $\gamma >0$  and $[\gamma x/(1-\alpha)] + \beta >0$.
In general, we expect  $\alpha<1$ since $\alpha>1$ would correspond to  implausible Increasing ARA (IARA) while $\beta \in \Real$. The family of HARA utility functions has three parameters $(\alpha,\beta,\gamma)$, whereas heterogeneity in risk aversion is one-dimensional in our setting.  We can achieve the latter  by fixing two  of these three utility parameters, thereby leaving the third one to describe individuals' heterogeneity in risk aversion. For instance, the CARA specification  $U(x;a)= - \exp(- ax)$ corresponds to $\beta=  [1- (1/\alpha)]^{-1/\alpha}$ and  $\gamma= a [1-(1/\alpha)]^{1-(1/\alpha)}$  with $\alpha  \rightarrow \pm \infty$.  See also  Lemma C.4 in Appendix C.
Alternatively, the CRRA specification $U(x;a)= x^{1-a}/(1-a)$ corresponds to $\beta=0$ and $\gamma=(1-\alpha)^{1-(1/\alpha)}$. A strictly negative value of $\beta$ implies Decreasing RRA  (DRRA) preferences, while a strictly  positive value of $\beta$ implies Increasing RRA (IRRA) preferences. 
These specifications satisfy A1'-(i) and  allow the nature of risk aversion $a$ to vary across individuals.  
Regarding A1'-(iii), several families of distributions satisfy it such as  the Poisson distribution ${\cal P}(\cdot;\theta)$ for $\theta>0$,    
the Negative Binomial distribution ${\cal NB}(\theta,p)$ for  $\theta>0$ and fixed $p\in(0,1)$,  and the Binomial distribution ${\cal B}(n,\theta)$ for $\theta\in(0,1)$ and fixed $n\geq 1$.  See Lemma  C.2 in Appendix C.  

We can now define the certainty equivalent $CE(t,dd;\theta,a,w)$ for a $(\theta,a)$-individual with wealth $w$. Following (1) and (2), $CE(t,dd;\theta,a,w)$ solves
\begin{eqnarray}
U(CE;a)  \equiv {\rm E} \left[U\Big(w-t - \sum_{j=0}^J \min\{dd,D_j\};a\Big)|\theta\right],
\end{eqnarray} 
where $J \sim P(\cdot;\theta)$, $D_j \stackrel{iid}{\sim} H(\cdot)$ and $D_0\equiv 0$ by convention.  Thus the frontier $\theta_{c,c+1}(\cdot)$ between coverages $(t_c,dd_c)$ and $(t_{c+1},dd_{c+1})$ satisfies 
\begin{eqnarray}
CE(t_{c},dd_{c};\theta_{c,c+1}(a),a,w) -  CE(t_{c+1},dd_{c+1};\theta_{c,c+1}(a),a,w) = 0
\end{eqnarray}
for all $a\in[\underline{a},\overline{a}]$. The frontier $\theta_{c,c+1}(\cdot)$ actually depends on wealth $w$. To simplify the notation,  we omit this dependence.  We also let $CE_c(\theta,a) \equiv CE(t_c,dd_c;\theta,a,w)$. 

The next lemma  extends Lemma 1 while allowing for  more than two contracts.  

\medskip\noindent
{\bf Lemma 1':} {\em Let A1' hold. Let the coverages $(t_c,dd_c)$, $c=1,\ldots, C \geq 2$ satisfy the RP condition (7). \\
(i) When $dd_c=0$ (full coverage), $CE_c(\theta,a)$ reduces to $w-t_c$. When $dd_c>0$, $CE_c(\theta,a)$  decreases in both risk $\theta$ and risk aversion $a$. \\
(ii) Suppose in addition that $CE_c(\theta,a)$ is supermodular in $(c,\theta)$ and $(c,a)$.\footnote{A function $\psi(c,x,y)$ is supermodular in $(c,x)$ whenever $\psi(c',x',y)+\psi(c,x,y) > \psi(c',x,y) + \psi(c,x',y)$ for all $c'>c$, $x'>x$ and $y$.  If $\psi(c,x,y)$ is differentiable in $x$ this is equivalent to $\partial \psi(c,x,y)/\partial x$ increasing in $c$ for all $(c,x,y)$, or $\partial \psi(c',x,y)/\partial x - \partial \psi(c,x,y)/\partial x >0$ for all $c'>c$ and $(x,y)$. If $\psi(c,x,y)$ was differentiable in $(c,x)$, this would be also equivalent to $\partial^2 \psi(c,x,y)/\partial c\partial x > 0$. See Topkis (1978).}  
Conditional on wealth $w$, the frontier $\theta_{c,c+1}(\cdot)$ between the coverages $(t_c,dd_c)$ and $(t_{c+1},dd_{c+1})$ is decreasing in the $(\theta,a)$-space.  Every $(\theta,a)$-individual  below (resp. above) this frontier prefers coverage $c$ over coverage $c+1$ (resp. $c+1$ over $c$). 
}

\medskip\noindent
{\bf Proof:} 
(i) When $dd=0$, (S.6) reduces to $U(CE;a)={\rm E}[U(w-t;a)]=U[w-t;a)]$.  Thus $CE(t,dd;\theta,a,w)=w-t$ since $U(\cdot;a)$ is increasing by A1'-(i).
Next, consider the case  $dd>0$.  Let $J' \sim P(\cdot;\theta')$ and $J \sim P(\cdot;\theta)$ where $\theta'>\theta$ so that $P(\cdot;\theta')\stackrel{FOSD}{\succ} P(\cdot;\theta)$ by A1'-(iii).  Setting $X_j\equiv \min\{dd,D_j\}$ for $j\in\{0,1,2,\ldots\}$ in Lemma C.6, we have $\sum_{j=0}^{J'} \min\{dd,D_j\} \stackrel{FOSD}{\succ} \sum_{j=0}^J \min\{dd,D_j\}$.    Hence $w-t- \sum_{j=0}^{J} \min\{dd,D_j\} \stackrel{FOSD}{\succ} w-t- \sum_{j=0}^{J'} \min\{dd,D_j\}$.  Because $U(\cdot;a)$ is increasing by A1'-(i), it follows from e.g. Gollier (2001) that 
${\rm E}\left[U(w-t- \sum_{j=0}^{J} \min\{dd,D_j\};a)\right] > {\rm E}\left[U(w-t- \sum_{j=0}^{J'} \min\{dd,D_j\};a)\right] $.
That is, the certainty equivalent $CE(t,dd;$ $\theta,a,w)$ defined by (S.6) is decreasing in $\theta$.
It remains to verify that $CE(t,dd; \theta,a,w)$ is decreasing in $a$.  Indeed because the $a'$-individual is more risk averse than the $a$-individual when $a'>a$ by A1'-(i), it follows that the former is worse off than the latter when offered the same lottery $w-t- \sum_{j=0}^{J} \min\{dd,D_j\}$.  See Rothschild  and Stiglitz (1970) or e.g. Gollier (2001).  

(ii)   The frontier between the two coverages $(t_c,dd_c)$ and $(t_{c+1},dd_{c+1})$ is the locus of $(\theta,a)$ pairs satisfying the indifference condition  (S.7), i.e., 
\begin{eqnarray}
CE_{c}(\theta,a) -  CE_{c+1}(\theta,a) = 0.
\end{eqnarray} 
Let $(\theta,a)$ and $(\theta',a')$ be two points on this locus with $a'>a$.  We want to show that $\theta' < \theta$.  Because $CE_c(\theta,a)$ is supermodular in $(c,a)$, we have $CE_c(\theta,a')-CE_c(\theta,a)<CE_{c+1}(\theta,a')-CE_{c+1}(\theta,a)$ or upon rearranging terms
\begin{eqnarray}
CE_c(\theta,a')-CE_{c+1}(\theta,a')<CE_c(\theta,a)-CE_{c+1}(\theta,a) = 0
\end{eqnarray}
using (S.8).  In particular, (S.9) shows that $\theta'\neq\theta$ since $(\theta',a')$ being on the indifference locus satisfies $CE_c(\theta',a')-CE_{c+1}(\theta',a')=0$.  Moreover, subtracting the latter equation from (S.9) gives upon rearranging terms
\begin{eqnarray*}
CE_{c+1}(\theta',a') - CE_{c+1}(\theta,a') 
< CE_c(\theta',a') - CE_c(\theta,a').
\end{eqnarray*}
Because $CE_c(\theta,a)$ is supermodular in $(c,\theta)$, we would obtain the reverse inequality if $\theta'>\theta$.
Thus $\theta'\leq\theta$ and hence $\theta'<\theta$ since $\theta'\neq\theta$ as shown earlier.

It remains to show the second part of (ii).  It suffices to show that any individual above or equivalently to the right of  the indifference locus prefers coverage ${c+1}$.  Let $(\theta',a')$ be such an individual and $(\theta',a)$ be the corresponding pair on the indifference locus with $a<a'$.  Similarly to (S.9) with $\theta$ replaced by $\theta'$ we obtain $CE_c(\theta',a')-CE_{c+1}(\theta',a')<0$. Thus the $(\theta',a')$-individual prefers coverage 2 as desired.\footnote{An alternative proof uses the differentiability of $CE_c(\theta,a)$ in $(\theta,a)$ for $c=1,2$. Specifically, the first part of (ii) can be proved by totally differentiating (A.3) to obtain the derivative $\theta'(a)$ of the frontier and using the differentiable version of the supermodularity of $CE_c(\theta,a)$ in $(c,\theta)$ and $(c,a)$.  To prove the second part of (ii), we can use
$CE_c(\theta',a') -  CE_c(\theta',a) = \int_{a}^{a'} \partial CE_c(\theta',\tilde{a})/\partial a \ d\tilde{a}$ where $\partial CE_c(\theta',\tilde{a})/\partial a$ is increasing in $c$.}
$\Box$

\medskip
Part (i)  of Lemma 1' conforms with intuition.  If an individual has a larger risk $\theta$, he/she will likely be involved in more accidents that will reduce his/her net wealth and thus his/her certainty equivalent.   Likewise, an individual with a higher risk aversion $a$ will have a lower utility and hence lower certainty equivalent for the coverage $(t,dd)$ {\em ceteris paribus}.  Regarding (ii),
supermodularity captures the idea that a better coverage is more valuable to insurees with higher $\theta$ (resp. higher $a$) than to insurees with lower $\theta$ (resp. lower $a$). Supermodularity conditions are widely used in economic theory.  See Milgrom and Roberts (1990), Vives (1990), and Athey (2002) among others.  In our case, this condition involves the terms of the coverages,  the utility function, the distribution of the number of accidents, and the distribution of damages since the expectation in (S.6) is taken with respect to $(J,D_1,\ldots,D_J)$.\footnote{As a matter of fact,  Lemma C.3 in Appendix C establishes the  supermodularity of $CE_c(\theta,a)$ in the CARA-Poisson specification of Section 2 irrespective of the damage distribution.  When the utility function is HARA, the difficulty arises from the non-explicit form of the certainty equivalent even when the distribution of the number of accidents is Poisson.  See Lemma C.4.}

The next lemma extends Lemma 2 by relaxing the CARA-Poisson assumptions.  As in Lemma 2, it ensures that the $C$ frontiers $\theta_{c,c+1}(\cdot)$ do not cross and lie on top of each other as $c$ increases from  1 to $C-1$.

\medskip\noindent
{\bf Lemma 2':} {\em Let A1' hold. Let the coverages $(t_c,dd_c)$,  $c=1,\ldots, C\geq 2$ satisfy the RP condition (7), and $CE_c(\theta,a)$ be supermodular in $(c,\theta)$ and $(c,a)$.  Suppose that
\begin{eqnarray}
\frac{ \frac{\partial CE_c(\theta_{c+1,c+2}(a),a)}{\partial a}
- \frac{\partial CE_{c+1}(\theta_{c+1,c+2}(a),a)}{\partial a} }
{ \frac{\partial CE_{c+1}(\theta_{c+1,c+2}(a),a)}{\partial a}
- \frac{\partial CE_{c+2}(\theta_{c+1,c+2}(a),a)}{\partial a} }
> 
\frac{ \frac{\partial CE_c(\theta_{c+1,c+2}(a),a)}{\partial \theta}
- \frac{\partial CE_{c+1}(\theta_{c+1,c+2}(a),a)}{\partial \theta} }
{ \frac{\partial CE_{c+1}(\theta_{c+1,c+2}(a),a)}{\partial \theta}
- \frac{\partial CE_{c+2}(\theta_{c+1,c+2}(a),a)}{\partial \theta} }
\end{eqnarray}
for $a\in[\underline{a},\overline{a}]$ and $c=1,\ldots,C-2$.
Then the frontiers $\theta_{c,c+1}(\cdot)$ between coverages $(t_c,dd_c)$ and $(t_{c+1},$ $d_{c+1})$ for $c=1,\dots,C-1$ satisfy
$\theta_{1,2}(\cdot)<\ldots<\theta_{C-1,C}(\cdot)$
on $[\underline{a},\overline{a}]$ if and only if 
\begin{eqnarray}
CE_c(\theta_{c+1,c+2}(\underline{a}),\underline{a}) < CE_{c+1}(\theta_{c+1,c+2}(\underline{a}),\underline{a})
\end{eqnarray}
for $c=1,\ldots,C-2$. 
}

\medskip\noindent
{\bf Proof:}  Fix $c=1,\ldots,C-2$.  We want to show that $\theta_{c,c+1}(\cdot)<\theta_{c+1,c+2}(\cdot)$ on $[\underline{a},\overline{a}]$.  From (S.7)  $\theta_{c,c+1}(\cdot)$ satisfies 
\begin{eqnarray}
CE_c(\theta_{c,c+1}(a),a) - CE_{c+1}(\theta_{c,c+1}(a),a)=0.
\end{eqnarray}
Because $CE_c(\theta,a)$ is supermodular in $(c,\theta)$, the derivative $\partial CE_c(\theta,a)/\partial\theta$ is increasing in $c$.  Thus $\partial CE_c(\theta,a)/\partial\theta - \partial CE_{c+1}(\theta,a)/\partial\theta<0$.  Hence $CE_c(\cdot,a) - CE_{c+1}(\cdot,a)$ is decreasing.  Thus, using (S.12), we have $\theta_{c,c+1}(a) < \theta_{c+1,c+2}(a)$ if and only if
\begin{eqnarray}
CE_c(\theta_{c+1,c+2}(a),a) - CE_{c+1}(\theta_{c+1,c+2}(a),a)<0.
\end{eqnarray}

We show that the LHS of (S.13) decreases in $a$ under condition (S.10).  Its (total) derivative  with respect to $a$ is
\begin{eqnarray}
\lefteqn{\frac{d}{d a}\Big[CE_c(\theta_{c+1,c+2}(a),a) - CE_{c+1}(\theta_{c+1,c+2}(a),a)\Big]}  \nonumber \\
&\qquad\qquad =& \left[ \frac{\partial CE_c(\theta_{c+1,c+2}(a),a)}{\partial\theta} - \frac{\partial CE_{c+1}(\theta_{c+1,c+2}(a),a)}{\partial\theta} \right] \theta'_{c+1,c+2}(a) \nonumber \\
&& + \left[ \frac{\partial CE_c(\theta_{c+1,c+2}(a),a)}{\partial a} - \frac{\partial CE_{c+1}(\theta_{c+1,c+2}(a),a)}{\partial a} \right],
\end{eqnarray}
where $\theta'_{c+1,c+2}(\cdot)$ is the derivative of the frontier $\theta_{c+1,c+2}(\cdot)$.  Because this frontier satisfies
$CE_{c+1}(\theta_{c+1,c+2}(a),a) - CE_{c+2}(\theta_{c+1,c+2}(a),a)=0$ by definition, differentiation with respect to $a$ gives 
\begin{eqnarray*}
\theta'_{c+1,c+2}(a) = - 
\frac{ \frac{\partial CE_{c+1}(\theta_{c+1,c+2}(a),a)}{\partial a} - \frac{\partial CE_{c+2}(\theta_{c+1,c+2}(a),a)}{\partial a} }
{ \frac{\partial CE_{c+1}(\theta_{c+1,c+2}(a),a)}{\partial\theta} - \frac{\partial CE_{c+2}(\theta_{c+1,c+2}(a),a)}{\partial\theta} } .
\end{eqnarray*}
Hence (S.14) gives upon rearranging terms
\begin{eqnarray*}
\lefteqn{ \frac{d}{d a}\Big[CE_c(\theta_{c+1,c+2}(a),a) - CE_{c+1}(\theta_{c+1,c+2}(a),a)\Big] } \\
&=& \!\!\! - \left[\frac{\partial CE_{c+1}(\theta_{c+1,c+2}(a),a)}{\partial a} - \frac{\partial CE_{c+2}(\theta_{c+1,c+2}(a),a)}{\partial a} \right] \\
&& \!\!\!\times \left[ \frac{ \frac{\partial CE_c(\theta_{c+1,c+2}(a),a)}{\partial\theta} - \frac{\partial CE_{c+1}(\theta_{c+1,c+2}(a),a)}{\partial\theta} }
{ \frac{\partial CE_{c+1}(\theta_{c+1,c+2}(a),a)}{\partial\theta} - \frac{\partial CE_{c+2}(\theta_{c+1,c+2}(a),a)}{\partial\theta} }
- \frac{ \frac{\partial CE_c(\theta_{c+1,c+2}(a),a)}{\partial a} - \frac{\partial CE_{c+1}(\theta_{c+1,c+2}(a),a)}{\partial a} }
{ \frac{\partial CE_{c+1}(\theta_{c+1,c+2}(a),a)}{\partial a} - \frac{\partial CE_{c+2}(\theta_{c+1,c+2}(a),a)}{\partial a} }
\right] 
\end{eqnarray*}
The term within the first pair of brackets is negative since $CE_c(\theta,a)$ is supermodular in $(c,a)$, so that $\partial CE_c(\theta,a)/\partial a$ is increasing in $c$.  The term within the second pair of brackets is negative by condition (S.10).  Thus the derivative of the LHS of (S.13) with respect to $a$ is negative as desired.
Because $CE_c(\theta_{c+1,c+2}(a),a) - CE_{c+1}(\theta_{c+1,c+2}(a),a)$ is decreasing in $a$, (S.13) holds for all $a\in[\underline{a},\overline{a}]$ if and only if it holds at $\underline{a}$.
$\Box$\

\medskip
Because $CE_c(\theta,a)$ is supermodular in $(c,\theta)$ and $(c,a)$,  numerators and denominators in (S.10) are negative.  Note that (S.10) is evaluated at the frontier $\theta_{c+1,c+2}(\cdot)$. Thus, a sufficient condition for (S.10) is 
\begin{eqnarray}
\frac{ \frac{\partial CE_c(\theta,a)}{\partial a}
- \frac{\partial CE_{c+1}(\theta,a)}{\partial a} }
{ \frac{\partial CE_c(\theta,a)}{\partial \theta}
- \frac{\partial CE_{c+1}(\theta,a)}{\partial \theta} }
> 
\frac{ \frac{\partial CE_{c+1}(\theta,a)}{\partial a}
- \frac{\partial CE_{c+2}(\theta,a)}{\partial a} }
{ \frac{\partial CE_{c+1}(\theta,a)}{\partial \theta}
- \frac{\partial CE_{c+2}(\theta,a)}{\partial \theta} }
\end{eqnarray}
for $(\theta,a)\in[\underline{\theta},\overline{\theta}]\times[\underline{a},\overline{a}]$ and $c=1,\ldots,C-2$.\footnote{Condition  (S.15) and hence  (S.10) are satisfied by the CARA-Poisson specification of Section 2 irrespective of the damage distribution.  See Lemma  C.5 in Appendix C.} Moreover, 
it follows from (S.7) that the LHS and RHS of (S.15) are the slopes  in absolute values of the frontiers $\theta_{c,c+1}(\cdot)$ and $\theta_{c+1,c+2}(\cdot)$, respectively, at $(\theta,a)$.   If these frontiers intersect  at $(\theta,a)$, (S.15) implies that they would  intersect  at most once.
As a matter of fact, under  (S.10), condition (S.11)  ensures that the $C$ frontiers $\theta_{c,c+1}(\cdot)$, $c=1,\ldots,C-1$ do not intersect.

Condition (S.11) generalizes condition (8)  of Lemma 2 to the case when the utility function and the distribution of the number of accidents are no longer CARA and Poisson, respectively. Indeed, using (2) to evaluate the difference $CE_c(\theta,a)-CE_{c+1}(\theta,a)$ at $(\theta_{c+1,c+2}(\underline{a}),\underline{a})$, where
the frontier $\theta_{c+1,c+2}(\cdot)$ is given by (3), it is easy to verify that condition (S.11) reduces to condition (8).  Moreover, from (S.6), condition (S.11) is equivalent to 
\begin{eqnarray*}
{\rm E}\! \left[U\Big(w \!-\! t_c \!-\!\!\! \sum_{j=0}^{J_{c+1,c+2}(\underline{a})} \!\!\!\!\min\{dd_c,D_j\};\underline{a}\Big) \right]
\!<\! {\rm E} \!\left[U\Big(w \!-\! t_{c+1} \!-\!\!\!\!\! \sum_{j=0}^{J_{c+1,c+2}(\underline{a})}\! \!\!\!\min\{dd_{c+1},D_j\};\underline{a}\Big) \right]
\end{eqnarray*}
for $c=1,\ldots,C-2$, where $J_{c+1,c+2}(\underline{a}) \sim P(\cdot;\theta_{c+1,c+2}(\underline{a}))$.  Since $U(\cdot;\underline{a})$ is increasing and concave, it follows  that a sufficient condition for (S.11) is 
\begin{eqnarray}
- t_c -\! \sum_{j=0}^{J_{c+1,c+2}(\underline{a})} \!\min\{dd_c,D_j\}
\stackrel{SOSD}{\prec}
- t_{c+1} - \! \sum_{j=0}^{J_{c+1,c+2}(\underline{a})} \!\min\{dd_{c+1},D_j\}
\end{eqnarray}
by definition of  Second-Order Stochastic Dominance (SOSD).
See Rothschild and Stiglitz (1970), or e.g., Gollier (2001). In particular, (S.16) is independent of wealth $w$.  Furthermore, upon taking expectations and invoking A1-(iv), (S.16) requires that $t_{c+1}-t_c < {\rm E}[J_{c+1,c+2}(\underline{a})] {\rm E}\Big[\min\{dd_c,D\}-\min\{dd_{c+1},D\}\Big]$, i.e., 
\begin{eqnarray}
t_{c+1}-t_c < {\rm E}[J_{c+1,c+2}(\underline{a})] \int_{dd_{c+1}}^{dd_c} [1-H(D)]dD,
\end{eqnarray}
where we have used the identity $\min\{dd_c,D\}=\min\{dd_{c+1},D\}$ $+ \left[ \min\{dd_c,D\} - dd_{c+1}\right]$ $ \Unit(dd_{c+1}\leq D)$ since $dd_c > dd_{c+1}$.  Because the expectation ${\rm E}[J_{c+1,c+2}(\underline{a})]$ depends on the distribution of damage $H(\cdot)$, the lowest risk aversion $\underline{a}$ as well as the terms $(t_{c+1},dd_{c+1})$ and $(t_{c+2},dd_{c+2})$ of the coverages $c+1$ and $c+2$, (S.17) is related to reverse nonlinear pricing as  for the CARA-Poisson case.\footnote{In the latter case, ${\rm E}[J_{c+1,c+2}(\underline{a})]=\theta_{c+1,c+2}(\underline{a})$.  From (3),  it follows that (S.17)   reduces to   condition (9) by letting $\underline{a}$ approaching zero.}

Lastly, we show how to extend the identification results of Section 3.
In line with A2, we make the following assumption.

\medskip\noindent
{\bf Assumption A2':} {\em While A2-(ii,iv) remain, A2-(i,iii)  are replaced by
A1'-(i,iii). In addition, the  family of distributions $P(\cdot;\theta)$ is additively closed.\footnote{A family of distributions $\{P(\cdot;\theta);\theta\in\Real\}$ is additively closed if their characteristic functions $\phi(\cdot;\theta)$ satisfy $\phi(\cdot;\theta_1)\phi(\cdot;\theta_2)=\phi(\cdot;\theta_1+\theta_2)$ for every $(\theta_1,\theta_2)\in\Real^2$.  This is equivalent to $\phi(\cdot;\theta)=\phi(\cdot;1)^\theta$.  See Rao (1992).}
}

\medskip\noindent
The distribution of accidents given $X$ is a mixture of the distribution of the number $J$ of accidents given $\theta$ with mixing distribution given by $F_{\theta|X}(\cdot|\cdot)$, i.e., $\Pr[J\leq\cdot|X]=\int P(\cdot|\theta) dF(\theta|X)$ using A2'-(iii).  The additive closedness  of $P(\cdot;\theta)$ then ensures the identification of the distribution $F_{\theta|X}(\cdot|\cdot)$ from the distribution of the number $J$ of accidents given $X$. See Teicher (1961) and Rao (1992).
Several families of discrete distributions are additively closed.  For instance, the Poisson distribution ${\cal P}(\theta)$ for $\theta>0$ and    
the Negative Binomial distribution ${\cal NB}(r,p)$ for  $r>0$ and fixed $p\in(0,1)$ mentioned earlier are additively closed.  In contrast, the Binomial distribution ${\cal B}(n,p)$ for $p\in(0,1)$ and fixed $n\geq 1$ is not.\footnote{When $n=1$, i.e., when one only observes whether an individual has an accident or not, 
Aryal, Perrigne and  Vuong (2012) show that our insurance model with CARA utility function is not identified despite exploiting all the restrictions of the model.} Given the identification of $F_{\theta|X}(\cdot|\cdot)$ under A2',  we use  steps 2 and 3 of Section 3.\footnote{The car value as a proxy for wealth $w$  is included in $X$.}
Hence the joint distribution $F(\theta,a|X)$ of risk and risk aversion is identified under the exclusion and support assumption A3.  As noted above,   the latter can be weakened as it suffices that the combined variations of the $C-1$ frontiers span the $\Theta(X) \times {\cal A}(X_0)$ space.
\vspace{0.8cm}

\begin{center}
{\bf Appendix C}
\end{center}
\medskip\noindent
This appendix establishes several lemmas mentioned in the text or used in Appendix B above.  Some  can be of independent interest. Others are likely known though we have yet found references for them.

\medskip\noindent
{\bf Lemma C.1:} {\em Let $H(\cdot)$ be a distribution function with support $(0,\overline{y})$. For any $y_\dagger\in(0,\overline{y})$, the function $\psi(x,y)\equiv\int_{y_\dagger}^y e^{xz}[1-H(z)] dz$ is log-supermodular in $(x,y)\in(0,+\infty)\times(y_\dagger,\overline{y})$. 
}

\medskip\noindent
{\bf Proof of Lemma C.1:}  Following Topkis (1978), the function $\psi(x,y)$ is log-supermodular if $\partial^2 \log \psi(x,y)/ \partial x\partial y >0$.    Taking derivatives, we obtain
\begin{eqnarray*}
\frac{\partial\log \psi(x,y)}{\partial y} &=& 
\frac{e^{xy}[1-H(y)]}{\psi(x,y)} \\
\frac{\partial^2\log \psi(x,y)}{\partial x \partial y} &=& 
\frac{ye^{xy}[1-H(y)]\psi(x,y)-e^{xy}[1-H(y)]\int_{y_\dagger}^y ze^{xz}[1-H(z)] \ dz}{\psi(x,y)^2}. 
\end{eqnarray*}
But  $\int_{y_\dagger}^yze^{xz}[1-H(z)] \ dz < y \int_{y_\dagger}^ye^{xz}[1-H(z)] \ dz = y \psi(x,y)$ since $z\leq y$ and $y_\dagger<y$.  Thus the numerator of $\partial^2\log \psi(x,y)/\partial x \partial y$ is positive as desired. $\Box$

\medskip\noindent
{\bf Lemma C.2:} {\em The Poisson distribution ${\cal P}(\theta)$ for $\theta>0$,    
the Negative Binomial distribution ${\cal NB}(r,p)$ for  $r>0$ and fixed $p\in(0,1)$,  and the Binomial distribution ${\cal B}(n,p)$ for $p\in(0,1)$ and fixed $n\geq 1$ are families of increasingly FOSD distributions as $\theta$, $r$ and $p$ increase respectively.
}

\medskip\noindent
{\bf Proof of Lemma C.2:}
From e.g., Johnson, Kemp and Kotz (2005), 
the Poisson cdf  with parameter $\theta$ is given by $\Pr[X\leq j]=  1-\Pr[{\rm Gamma}(j+1) \leq \theta]$
which is decreasing in $\theta$ for every $j=0,1,\ldots$.  Thus ${\cal P}(\theta')\stackrel{FOSD}{\succ}{\cal P}(\theta)$ whenever $\theta'>\theta$.  Similarly, 
the cdf of ${\cal NB}(r,p)$ is given by $\Pr[X\leq j]= \Pr[{\rm Beta}(r,j+1)\leq p]$, which is decreasing in $r$ for every $j=0,1,\ldots$.   
To see the latter, we note that  
${\rm Beta}(a,b) \stackrel{d}{=} 1-{\rm Gamma}(b)/[{\rm Gamma}(a)+{\rm Gamma}(b)]$ where ${\rm Gamma}(a)$ is independent of ${\rm Gamma}(b)$. Since ${\rm Gamma}(a') \stackrel{d}{=}{\rm Gamma}(a)+{\rm Gamma}(a'-a)$ where ${\rm Gamma}(a)$ is independent of ${\rm Gamma}(a'-a) \geq 0$, it follows that
${\rm Gamma}(a')\stackrel{FOSD}{\succ}{\rm Gamma}(a)$ for $a'>a$ implying that $\Pr[{\rm Beta}(r',j+1)\leq p] < \Pr[{\rm Beta}(r,j+1)\leq p]$ for $r'>r$.  
That is, ${\cal NB}(r',p)\stackrel{FOSD}{\succ}{\cal NB}(r,p)$ whenever $r'>r$. 
Lastly, the cdf of 
${\cal B}(n,p)$ is given by $\Pr[X\leq j] = \Pr[{\rm Beta}(n-j,j+1) \leq 1-p]$
which is decreasing in $p$ for every $j=0,1,\ldots,n$.  Thus ${\cal B}(n,p') \stackrel{FOSD}{\succ}{\cal B}(n,p)$ whenever $p'>p$.
$\Box$

\medskip\noindent
{\bf Lemma C.3:} {\em Let the coverages $(t_c,dd_c)$, $c=0,1,\ldots,C\geq 2$, satisfy the RP condition (7). The certainty equivalent $CE_c(\theta,a)$ of contract $c$ is supermodular in $(c,\theta)$ and $(c,a)$ when $U(\cdot;a)$ and $P(\cdot;\theta)$ are the CARA utility and the Poisson distribution, respectively.
}

\medskip\noindent
{\bf Proof of Lemma C.3:} We want to show that 
\begin{eqnarray*}
\frac{\partial  CE_{c'}(\theta,a)}{\partial \theta}
- \frac{\partial  CE_c(\theta,a)}{\partial \theta} >0
\quad {\rm and} \quad
\frac{\partial  CE_{c'}(\theta,a)}{\partial a}
- \frac{\partial  CE_c(\theta,a)}{\partial a} >0
\end{eqnarray*}
for $c<c'$. To prove the first inequality, we note that (2) gives
\begin{eqnarray*}
\frac{\partial  CE_{c'}(\theta,a)}{\partial \theta}
- \frac{\partial  CE_c(\theta,a)}{\partial \theta} 
= -\frac{1}{a}[\phi_a(dd')-\phi_a(dd)] .
\end{eqnarray*}
Because $\phi_a(\cdot)$ is increasing and $dd'<dd$, the desired inequality follows.  Regarding the second inequality, we show that $\partial  CE_c(\theta,a)/\partial a$ increases in $c$ or equivalently decreases in $dd$.  From (2) this is equivalent to showing that
\begin{eqnarray*}
\frac{\partial  CE_c(\theta,a)}{\partial a} = 
- \theta \frac{\partial}{\partial a}\left[\frac{\phi_a(dd)-1}{a}\right]
\end{eqnarray*}
is decreasing in $dd$. But $[\phi_a(dd)-1]/a=\int_0^{dd}e^{aD}[1-H(D)]dD$ by (A.1).  Thus, 
\begin{eqnarray*}
\frac{\partial^2}{\partial dd \partial a}\left[\frac{\phi_a(dd)-1}{a}\right] 
= \frac{\partial}{\partial a} \left[e^{add}[1-H(dd)]\right]
= ae^{add}[1-H(dd)] > 0.
\end{eqnarray*}
Hence $\partial  CE_c(\theta,a)/\partial a$ is decreasing in $dd$ as desired. 
$\Box$

\medskip\noindent
{\bf Lemma C.4:} {\em Let $U(x; a, \phi, \delta) 
= - \delta( 1 - \phi ax )^{1/\phi}$ 
for $a>0$, $\phi<1$, $\delta>0$ and $1 - \phi ax > 0$. 
Then the following holds

(i) $U(\cdot; a, \phi, \delta)$ is a HARA utility.  The CARA utility is obtained as $(\phi, \delta) \rightarrow (0,1)$.

(ii) For any fixed $\phi<1$ and $\delta>0$, $U(x; a, \phi, \delta)$ satisfies Assumption 1'-(i). \\
Moreover, let Assumption A1'-(ii,iii,iv) holds with $P(\cdot;\theta)$ as the Poisson distribution and the coverages $(t_c,dd_c)$, $c=0,1,\ldots,C\geq 2$, satisfy the RP condition (7). 

(iii) The certainty equivalent $CE_c(\theta,a,\phi)$ of contract $c$ does not depend on $\delta$.  Furthermore,
$CE_c(\theta,a,\phi)$ is $\epsilon$-supermodular in $(c,\theta)$ and $(c,a)$ in the sense that for every $\epsilon>0$ there exists $\eta=\eta(\epsilon)>0$ such that 
\begin{eqnarray}
CE_{c'}(\theta',a,\phi) - CE_c(\theta',a,\phi) 
- CE_{c'}(\theta,a,\phi) + CE_c(\theta,a,\phi) &>& 0 
\\
CE_{c'}(\theta,a',\phi) - CE_c(\theta,a',\phi) 
- CE_{c'}(\theta,a,\phi) + CE_c(\theta,a,\phi) &>& 0
\end{eqnarray}
for all $c'>c$, $\theta' \geq \theta + \epsilon$, $a' \geq a + \epsilon$, $(\theta,a)\in[\underline{\theta},\overline{\theta}]\times[\underline{a},\overline{a}]$  and $|\phi| < \eta$ whenever $CE_c(\cdot,\cdot,\cdot)$ is continuous in $(\theta,a,\phi)$.\footnote{ Supermodularity in $(c,\theta)$ and $(c,a)$ replaces $\theta' \geq \theta + \epsilon$ and  $a' \geq a + \epsilon$ for any $\epsilon>0$ by  $\theta' > \theta$ and $a' > a$, respectively.  See footnote 5.} 
}

\medskip\noindent
{\bf Proof of Lemma C.4:} (i) As is well known, the CARA utility with parameter $a>0$ is a special case of the HARA utility by setting $\beta = [1- (1/\alpha)]^{-1/\alpha}$, $\gamma = a[1-(1/\alpha)]^{1-(1/\alpha)}$ and letting $\alpha \rightarrow \pm \infty$.  Indeed, upon substituting,  we obtain after some algebra 
\begin{eqnarray*}
U(x;\alpha,\beta,\gamma) = - \left[ 1 - \frac{ax}{\alpha} \right]^{\alpha}  \longrightarrow -\exp(-ax)
\end{eqnarray*}
as $\alpha \rightarrow\pm\infty$.  To avoid dealing with a divergence to infinity and  to obtain the CARA utility, we reparameterize the HARA utility function  when $1/\alpha < 1$ and $\beta>0$ by letting $\phi = 1/\alpha$, $a = \gamma/\{\beta[1- (1/\alpha)] \}$ and $\delta =[1- (1/\alpha)] \beta^{\alpha}$.  Hence
\begin{eqnarray*}
U(x;\alpha,\beta,\gamma) = U(x; a, \phi, \delta) 
= - \delta( 1 - \phi ax )^{1/\phi}
\end{eqnarray*}
for $a>0$, $\phi<1$, $\delta>0$ and $1 - \phi ax > 0$.  Because it is a HARA utility, $U(\cdot; a, \phi,\delta)$ is increasing and concave on $(-\infty,1/(\phi a))$ if $\phi>0$ and on $(1/(\phi a),+\infty)$ if $\phi<0$. 
In particular, the CARA utility with parameter $a$ is obtained  when $(\phi,\delta) \rightarrow (0,1)$.

(ii)  Fix $\phi<1$ and $\delta>0$.  Let $a'>a$.  We show that $U(x; a', \phi, \delta) = q[U(x; a, \phi, \delta)]$ for some increasing and concave function $q(\cdot)$.  See footnote 5.  We have
\begin{eqnarray*}
U(x; a', \phi, \delta) 
= - \delta\left( 1 - \frac{a'}{a} \phi ax \right)^{1/\phi}
= - \delta\left\{ 1 - \frac{a'}{a} \left[ 
1 - \left(\frac{-U(x; a, \phi, \delta)}{\delta} \right)^{\phi}
\right]\right\}^{1/\phi}
\end{eqnarray*}
using the definition of $U(x; a, \phi, \delta)$, which is negative.  Thus, 
\begin{eqnarray*}
q(u) = - \delta\left\{ 1 - \frac{a'}{a} \left[ 
1 - \left(\frac{-u}{\delta} \right)^{\phi}
\right]\right\}^{1/\phi}.
\end{eqnarray*}
Hence, after some algebra its first derivative is 
\begin{eqnarray*}
q'(u) = \frac{a'}{a} \left(\frac{-u}{\delta}\right)^{\phi-1}
\left\{ 1 - \frac{a'}{a} \left[ 
1 - \left(\frac{-u}{\delta} \right)^{\phi}
\right]\right\}^{(1/\phi)-1} > 0.
\end{eqnarray*}
Thus, $q(.)$ is increasing as desired.  Its second derivative is
\begin{eqnarray*}
q^{\prime\prime}(u) &=& \frac{a'}{a} 
\left(\frac{-u}{\delta}\right)^{\phi-2}
\left\{ 1 - \frac{a'}{a} \left[ 1 - \left(\frac{-u}{\delta} \right)^{\phi}
\right]\right\}^{(1/\phi)-2} \\
&& \times \left( - \frac{\phi-1}{\delta}
\left\{ 1 - \frac{a'}{a} \left[ 1 - \left(\frac{-u}{\delta} \right)^{\phi}
\right]\right\}
+ \left(\frac{1}{\phi}-1\right) \frac{a'}{a} \left(\frac{-\phi}{\delta}\right)
\left(\frac{-u}{\delta}\right)^{\phi} \right) \\
&=& \frac{1-\phi}{\delta} \left(1-\frac{a'}{a}\right)  \frac{a'}{a} 
\left(\frac{-u}{\delta}\right)^{\phi-2}
\left\{ 1 - \frac{a'}{a} \left[ 1 - \left(\frac{-u}{\delta} \right)^{\phi}
\right]\right\}^{(1/\phi)-2}  < 0 
\end{eqnarray*}
since $0<a<a'$, $\phi<1$ and $\delta>0$.  
Thus, $q(\cdot)$ is concave as desired.

(iii) It follows from (S.6) that the certainty equivalent $CE_c(\theta,a,\phi)$  of contract $c$ does not depend on $\delta$ when the utility is $U(x; a, \phi, \delta) 
= - \delta( 1 - \phi ax )^{1/\phi}$. 
Let $\Delta_{c'c}(\theta,a,\phi) \equiv CE_{c'}(\theta,a,\phi) - CE_c(\theta,a,\phi)$.   Thus (S.18) and (S.19) become
\begin{eqnarray*}
\Delta_{c'c}^2(\theta',\theta,a,\phi) &\equiv &
\Delta_{c'c}(\theta',a,\phi) - \Delta_{c'c}(\theta,a,\phi) > 0
\\
\Delta_{c'c}^2(\theta,a',a,\phi) &\equiv& 
\Delta_{c'c}(\theta,a',\phi) - \Delta_{c'c}(\theta,a,\phi) > 0.
\end{eqnarray*}
In particular, $CE_c(\theta,a,0)$ is the certainty equivalent of contract $c$ when $U(\cdot;a,\phi,\delta)$ is the CARA utility. It follows from Lemma C.3 that (S.18) and (S.19) hold for all $c'>c$, $\theta'>\theta$, $a'>a$ and $(\theta,a)\in[\underline{\theta},\overline{\theta}]\times[\underline{a},\overline{a}]$ when $\phi=0$.  

Consider (S.18).  For $\epsilon>0$ small, let ${\cal K}_{\epsilon} \equiv \{(\theta',\theta) \in [\underline{\theta},\overline{\theta}]^2: \theta' \geq \theta + \epsilon\}$.  Also, consider $[\underline{\phi},\overline{\phi}]$  with 
$-\infty<\underline{\phi}< 0 < \overline{\phi} <1$.
Fix $c' >c $. 
Because $CE_c(\theta,a,\phi)$ is continuous in $(\theta,a,\phi) \in [\underline{\theta},\overline{\theta}]\times[\underline{a},\overline{a}]\times[\underline{\phi},\overline{\phi}]$, then  
$\Delta_{c'c}^2(\theta',\theta,a,\phi)$ is continuous  and hence uniformly continuous in 
$(\theta',\theta,a,\phi) \in {\cal K}_{\epsilon}\times[\underline{a},\overline{a}]\times[\underline{\phi},\overline{\phi}]$.  Thus, for every $\tilde{\epsilon}>0$ there exists $\eta=\eta(\epsilon,\tilde{\epsilon})$ independent of $(\theta',\theta,a)$ such that 
\begin{eqnarray*}
|\phi| <  \eta \quad \Rightarrow \quad 
| \Delta_{c'c}^2(\theta',\theta,a,\phi) 
- \Delta_{c'c}^2(\theta',\theta,a,0) | < \tilde{\epsilon}
\end{eqnarray*}
for all $(\theta',\theta,a) \in {\cal K}_{\epsilon}\times[\underline{a},\overline{a}]$. 
But $\Delta_{c'c}^2(\theta',\theta,a,0)$ is positive and continuous in $(\theta',\theta,a) \in {\cal K}_{\epsilon}\times[\underline{a},\overline{a}]$.  Thus, $\Delta_{c'c}^2(\cdot,\cdot,\cdot,0)> \underline{m}$ on ${\cal K}_{\epsilon}\times[\underline{a},\overline{a}]$ for some $\underline{m}=\underline{m}(\epsilon)>0$.   Hence, letting $\tilde{\epsilon}= \underline{m}/2$ there exists $\eta=\eta(\epsilon)$ independent of $(\theta',\theta,a)$ such that 
\begin{eqnarray*}
\Delta_{c'c}^2(\theta',\theta,a,\phi) = 
\Delta_{c'c}^2(\theta',\theta,a,0) 
+ \left[ \Delta_{c'c}^2(\theta',\theta,a,\phi) 
- \Delta_{c'c}^2(\theta',\theta,a,0) \right] > \underline{m}/2 >0
\end{eqnarray*}
for all $(\theta',\theta,a) \in {\cal K}_{\epsilon}\times[\underline{a},\overline{a}]$ and $|\phi| <  \eta$. 
Because there is a finite number of pairs $(c,c')$ such that $c'>c$, this argument shows that (S.18) also holds for all $c' > c$.  A similar argument with ${\cal K}_{\epsilon} \equiv \{(a',a) \in [\underline{a},\overline{a}]^2: a' \geq a + \epsilon\}$ establishes (S.19). 
$\Box$

\medskip\noindent
{\bf Lemma C.5:} {\em Let the coverages $(t_c,dd_c)$, $c=0,1,\ldots,C\geq 2$ satisfy the RP condition (7). The certainty equivalent $CE_c(\theta,a)$ associated with coverage $c$ satisfies condition (S.10) when $U(\cdot;a)$ and $P(\cdot;\theta)$ are the CARA utility and the Poisson distribution, respectively.
}

\medskip\noindent
{\bf Proof of Lemma C.5:} From the proof of Lemma C.3, we have 
\begin{eqnarray*}
\frac{\partial CE_{c}(\theta,a)}{\partial \theta}
- \frac{\partial CE_{c+1}(\theta,a)}{\partial \theta}
&=& - \frac{1}{a}[\phi_a(dd_c)-\phi_a(dd_{c+1})] 
= - \int_{dd_{c+1}}^{dd_c}e^{aD}[1-H(D)]dD, \\
\frac{\partial CE_{c}(\theta,a)}{\partial a}
- \frac{\partial CE_{c+1}(\theta,a)}{\partial a} 
&=& - \theta \frac{\partial}{\partial a} \left[ \int_{dd_{c+1}}^{dd_c}e^{aD}[1-H(D)]dD \right] .
\end{eqnarray*}
Thus upon simplifying and rearraging terms, condition (S.10) is equivalent to
\begin{eqnarray*}
\frac{\partial}{\partial a}\left[ \log \int_{dd_{c+1}}^{dd_c}e^{aD}[1-H(D)]dD \right] >
\frac{\partial}{\partial a}\left[ \log \int_{dd_{c+2}}^{dd_{c+1}}e^{aD}[1-H(D)]dD \right],
\end{eqnarray*}
which is equivalent to 
\begin{eqnarray*}
\frac{\partial}{\partial a}\left[ \int_{dd_{c+1}}^{dd_c}e^{aD}[1-H(D)]dD / 
\int_{dd_{c+2}}^{dd_{c+1}}e^{aD}[1-H(D)]dD \right] >0,
\end{eqnarray*}
i.e., that the ratio within brackets is increasing in $a$.  Because
\begin{eqnarray*}
\frac{ \int_{dd_{c+2}}^{dd_c}e^{aD}[1-H(D)]dD } 
{ \int_{dd_{c+2}}^{dd_{c+1}}e^{aD}[1-H(D)]dD } = 1 + 
\frac{ \int_{dd_{c+1}}^{dd_c}e^{aD}[1-H(D)]dD } 
{ \int_{dd_{c+2}}^{dd_{c+1}}e^{aD}[1-H(D)]dD },
\end{eqnarray*}
it follows that condition (S.10) is equivalent to the LHS increasing in $a$.  This is true because $\int_{dd_{c+2}}^{dd}e^{aD}[1-H(D)]dD$ is log-supermodular in $(a,dd)$ by Lemma C.1 upon letting $x=a$, $y=dd$ and $y_\dagger =dd_{c+2}$.

\medskip\noindent
{\bf Lemma C.6:} {\em Let $J$ and $J'$ be random variables distributed as $P(\cdot)$ and $P'(\cdot)$ on $\Natural\equiv\{0,1,2,\ldots\}$.  
Let $X_0\equiv 0$ and  $X_1,X_2,\ldots$ be i.i.d. random variables independent of $(J,J')$ with support $(0,\overline{x})$, $0<\overline{x}\leq+\infty$. Suppose that $P'(\cdot) \stackrel{FOSD}{\succ} P(\cdot)$.  Then $\sum_{i=0}^{J'} X_i \stackrel{FOSD}{\succ}\sum_{i=0}^{J} X_i$.
}

\medskip\noindent
{\bf Proof of Lemma C.6:} By definition, $\sum_{i=0}^{J'} X_i \stackrel{FOSD}{\succ}\sum_{i=0}^{J} X_i$ is equivalent to 
\begin{eqnarray}
\Pr\left[\sum_{i=0}^{J} X_i \leq x \right]  \geq 
\Pr\left[\sum_{i=0}^{J'} X_i \leq x \right]   
\end{eqnarray} 
for all $x \geq 0$ with strict inequality for some $x>0$.  Let $F_{S_j}(\cdot)$ be the cdf of $S_j \equiv \sum_{i=0}^{j} X_i=\sum_{i=1}^{j} X_i$ for $j\in\Natural$.  Note that $F_{S_0}(x)=1$ for $x\geq 0$.  Thus, we have
\begin{eqnarray*}
\Pr\left[\sum_{i=0}^{J} X_i \leq x \right] 
= {\rm E}\left\{\Pr\left[ \sum_{i=0}^{J} X_i \leq x | J \right] \right\}
= P(0) + \sum_{j=1}^{+\infty} F_{S_j}(x) P(j)  
= \sum_{j=0}^{+\infty} F_{S_j}(x) P(j) 
\end{eqnarray*}
for $x \geq 0$.
A similar expression holds when $J$ is replaced by $J'$ with $P(j)$ replaced by $P'(j)$.    Thus (S.20) is equivalent to
\begin{eqnarray*}
\sum_{j=0}^{+\infty}  F_{S_j}(x)  P(j) \geq
\sum_{j=0}^{+\infty} F_{S_j}(x) P'(j). 
\end{eqnarray*}
Let $U(j;x) \equiv - F_{S_j}(x)$ for $j\in \Natural$ and $x \geq 0$. Hence we want to show that 
\begin{eqnarray}
{\rm E}[U(J';x)] \geq {\rm E}[U(J;x)]
\end{eqnarray}
for all $x \geq 0$ with strict inequality for some $x>0$ whenever $P'(\cdot) \stackrel{FOSD}{\succ} P(\cdot)$.  

To this end, we show that $U(\cdot;x)$ is nondecreasing on $\Natural$  and increasing on $(0,\overline{x})$.  To see this, we note that $F_{S_{j+1}}(\cdot) \stackrel{FOSD}{\succ}  F_{S_j}(\cdot)$ for $j\geq 0$ because $S_{j+1} = S_j + X_{j+1}$ with $X_{j+1} \geq 0$ and $\overline{x}> 0$.  Specifically, for $j\in\Natural$ and $x \geq 0$, we have $F_{S_j}(x) \geq F_{S_{j+1}}(x)$ with strict inequality for $\{j=0 \ {\rm and}\ x\in[0,\overline{x})\}$ or $\{j\geq 1 \ {\rm and}\  x\in(0,(j+1)\overline{x})\}$.  Hence $U(j;x)$ is nondecreasing in $j\in\Natural$ for every $x\geq 0$.  Moreover, it is easy to verify that  $U(\cdot;x)$ is increasing on $\Natural$ when $x\in(0,\overline{x})$. 
Now, because $P'(\cdot) \stackrel{FOSD}{\succ} P(\cdot)$, it follows that (S.21) holds for all $x \geq 0$ with strict inequality 
for $x\in(0,\overline{x})$. $\Box$

\begin{center}
{\bf References}
\end{center}

\smallskip\noindent
{\bf Aryal, Gaurab}, {\bf Isabelle Perrigne}, and {\bf Quang Vuong} (2012): ``Nonidentification of Insurance Models with Probability of Accidents,'' Unpublished manuscript, University of Virginia.

\smallskip\noindent
{\bf Athey, Susan} (2002): ``Monotone Comparative Statics under Uncertainty,'' {\em Quarterly Journal of Economics}, 117, 187-223.

\smallskip\noindent
{\bf Chernozhukov, Victor}, {\bf Han Hong} and {\bf Elie Tamer} (2007): ``Estimation and Confidence Regions for Parameter 
Sets in Econometric Models,'' {\em Econometrica}, 75, 1243-1284.

\smallskip\noindent
{\bf Cohen, Alma} and {\bf Liran Einav} (2007): ``Estimating Risk Preferences from Deductible Choice,'' 
{\em American Economic Review}, 97, 745-788.

\smallskip\noindent
{\bf Gollier, Christian} (2001): {\em The Economics of Risk and Time}, MIT Press: Cambridge.

\smallskip\noindent
{\bf Haile, Philip} and {\bf Elie Tamer} (2003): ``Inference with an Incomplete Model of English Auctions,''
{\em Journal of Political Economy}, 111, 1-51.

\smallskip\noindent
{\bf Johnson, Norman L.}, {\bf Adrienne W. Kemp} and {\bf Samuel Kotz} (2005): {\em Univariate Discrete Distributions}, 3rd Edition, New Jersey, USA: John Wiley and Sons.

\smallskip\noindent
{\bf Manski, Charles F.} and {\bf Elie Tamer} (2002): ``Inference on Regressions with Interval Data on a Regressor or 
Outcome,'' {\em Econometrica}, 70, 519-546.

\smallskip\noindent
{\bf Milgrom, Paul} and {\bf John Roberts} (1990): ``Rationalizability, Learning, and Equilibrium in Games with Strategic Complementarities,'' {\em Econometrica}, 58, 1255-1277.

\smallskip\noindent
{\bf Rao, B.L.S. Prakasa} (1992): {\em Identifiability in Stochastic Models: Characterization of Probability Distributions}, New York, USA: Academic Press.

\smallskip\noindent
{\bf Rothschild, Michael} and {\bf Joseph Stiglitz} (1970): ``Increasing Risk: I. A Definition,'' {\em Journal of Economic Theory}, 2, 225-243.

\smallskip\noindent
{\bf Teicher, Henry} (1961): ``Identifiability of Mixtures,'' {\em Annals of Mathematical Statistics}, 32, 244-248.

\smallskip\noindent
{\bf Topkis, Donald} (1978): ``Minimizing a Submodular Function on a Lattice,'' {\em Operation Research}, 26, 305-321.

\smallskip\noindent
{\bf Vives, Xavier} (1990): ``Nash Equilibrium with Strategic Complementarities,'' {\em Journal of Mathematical Economics}, 19, 305-321.

\end{document}